\documentclass{emulateapj}

\def\sun{\ifmmode\odot\else$\odot$\fi}

\shorttitle{The extreme star formation activity of Arp~299}
\shortauthors{A. Alonso-Herrero et al.}

\begin{document}

\title{The Extreme Star Formation Activity of Arp~299 Revealed by Spitzer IRS
  Spectral Mapping\altaffilmark{8}}

\author{Almudena Alonso-Herrero\altaffilmark{1,2}, 
George  H. Rieke\altaffilmark{2}, Luis Colina\altaffilmark{1},
Miguel Pereira-Santaella\altaffilmark{1}, 
Macarena Garc\'{\i}a-Mar\'{\i}n\altaffilmark{1,3},  
J.-D. T. Smith\altaffilmark{4},
Bernhard  Brandl\altaffilmark{5}, Vassilis 
Charmandaris\altaffilmark{6} and Lee Armus\altaffilmark{7}} 

\altaffiltext{1}{Departamento de Astrof\'{\i}sica Molecular 
e Infrarroja, Instituto de Estructura de la Materia, CSIC, 
E-28006 Madrid, Spain; E-mail: aalonso@damir.iem.csic.es}
\altaffiltext{2}{Steward Observatory, University of Arizona, Tucson,  AZ 85721}
\altaffiltext{3}{I. Physikalisches Institut, Universit\"at zu K\"oln, 50937
 K\"oln, Germany}
\altaffiltext{4}{Ritter Astrophysical Research Center, University of
  Toledo, Toledo, OH 43603}
 \altaffiltext{5}{Leiden Observatory, Leiden University, P.O. Box 9513,
  2300 RA Leiden, The Netherlands}
 \altaffiltext{6}{Department of Physics, University of Crete,
   GR-71003, Heraklion, Greece}
 \altaffiltext{7}{Spitzer Science Center, California Institute of Technology, Pasadena, CA 91125}
\altaffiltext{8}{Based on observations obtained with the Spitzer Space Telescope, which is operated by the Jet Propulsion Laboratory, California Institute of Technology, under NASA contract 1407}

\begin{abstract}
We present {\it Spitzer}/IRS spectral mapping observations of the luminous
infrared galaxy (LIRG) Arp~299 (IC~694 + NGC~3690) 
covering the central $\sim 45\arcsec \sim 9\,{\rm kpc}$. 
The integrated mid-IR  spectrum of 
Arp~299 is  similar to that of local starbursts 
despite  its strongly interacting nature and high infrared luminosity, 
$L_{\rm IR} \sim 6 \times 10^{11}\,{\rm L}_\odot$.
This is explained  because the star formation (probed by 
e.g. high [NeIII]$15.56\,\mu$m/[NeII]$12.81\,\mu$m line ratios) is spread 
across at least 6-8\,kpc. Moreover, a large fraction of this star
formation is taking place 
in young regions of moderate mid-IR optical
depths such as the C+C$^\prime$ complex in the overlap 
region between the two galaxies and
in H\,{\sc ii} regions in the disks of the galaxies. It is 
only source A, the nuclear region  
of IC~694, that shows the typical mid-IR
characteristics of ultraluminous infrared galaxies (ULIRGs, 
$L_{\rm IR} > 10^{12}\,{\rm L}_\odot$), that 
is, very compact (less than 1\,kpc) 
and dust-enshrouded star formation resulting in a  deep 
silicate feature and
moderate equivalent widths of the PAHs. The nuclear region of
NGC~3690, known  as source B1, hosts a low-luminosity 
AGN and is surrounded by regions of star formation. Although 
the high
excitation [NeV]$14.32\,\mu$m line typical of AGN is not detected in B1, its 
upper limit is consistent with the value expected from the
X-ray luminosity. The AGN emission is detected in the form of a strong
hot dust component that accounts for $80-90\%$ of the $6\,\mu$m luminosity
of B1. The similarity between the  Arp~299 integrated mid-IR spectrum 
and those of high-$z$ ULIRGs suggests that Arp~299    
may represent a local example, albeit with lower
IR luminosity and possibly higher metallicity, of the star-formation
processes occurring at high-$z$.

\end{abstract}


\section{Introduction}

Luminous Infrared (IR) Galaxies (LIRGs, $L_{\rm IR}= 10^{11}-10^{12}\,{\rm
  L}_\odot$, see Sanders \& Mirabel 1996 for a review) 
are an important cosmological class 
as they are the main contributors to the co-moving star formation 
rate density of the universe at $z\sim 1$ 
(Elbaz et al. 2002; Le Floc'h et al. 2005; P\'erez-Gonz\'alez et al. 2005; 
Caputi et al. 2007). At higher redshifts ($z\sim 2$) LIRGs and 
Ultra Luminous Infrared Galaxies (ULIRGs, $L_{\rm IR}= 10^{12}-10^{13}\,{\rm
  L}_\odot$) have comparable contributions.  Moreover, a large fraction of
  massive galaxies at high redshift are LIRGs and ULIRGs, indicating 
that a phase of intense star formation and/or AGN activity dominated the
  energy output of these massive galaxies and was more common in the past 
than now (Caputi et al. 2006). 

\begin{table*}

\caption{Log of the Spitzer/IRS Spectral Mapping Observations}

\begin{tabular}{lccccccccc}
\hline
\hline

Target & RA & Dec & IRS & Map & Plate  & PA 
 & Ramp & Number & Program \\
       &    &     & Module & Size & Scale & & Duration & Cycles & ID\\
(1) & (2) & (3) & (4) & (5) & (6) & (7) & (8) & (9) & (10) \\
 \hline
Arp~299  & 11h28m32.8s &+58d33m42s & SL1 + SL2 & $42 \times 19$ & 1.85 & 56.82
&  14s & 2 & 21\\
         &  &                      & LL1 + LL2 & $37 \times 8 $ & 5.08 &
$-26.9$ & 14s & 2& 21\\
IC~694   &11h28m33.7s  &+58d33m46s & SH        & $16 \times 10$ & 2.26 &
157.38 & 
30s & 2 & 30577\\
         & &                       & LH        & $5 \times 6$   & 4.46 & 72.58
& 60s & 4 & 30577\\
NGC~3690 &11h28m30.7s  &+58d33m43s & SH        & $16 \times 10$ & 2.26 &
157.31 &  30s & 2 & 30577\\
         &             &           & LH        & $5 \times 6$   & 4.46 & 72.51 
& 60s & 4 & 30577\\
\hline
\end{tabular}
\tablecomments{(1) Target; (2) and (3) RA Dec (J2000) of
the center of the final data cube (before rotation) as given in the headers; 
(4) IRS Module; (5) Original size of the final map in
pixel before rotation; (6) 
Plate scale of the map in arcsec/pixel; (7) Average slit orientation over the
DCEs in degrees; (8) Ramp duration; (9) Number of cycles; (10) 
{\it Spitzer} archive Program ID.}
\end{table*}

Recent results with the 
Infrared Spectrograph (IRS) instrument (Houck et al. 2004) on {\it
  Spitzer} are now showing that the mid-IR spectra 
  of $z\sim 1.7- 2$ star-forming galaxies with high 
IR luminosities (in the ULIRG class, or even more luminous) appear to
  be more consistent with those of local starbursts and LIRGs, rather than
 local ULIRGs (Rigby et al. 2008; Farrah et al. 2008). 
In particular these high-$z$ IR galaxies have similar PAH features to
those of local ULIRGs, but  
substantially shallower $9.7\,\mu$m absorption features.
This may be due to lower optical depths, perhaps because star formation in
ULIRGs was  more spatially extended at high-$z$ than now, and/or
metallicity effects. 
Therefore, local LIRGs play a fundamental role for understanding their
  more distant and luminous counterparts. To  characterize fully the mid-IR
  properties of local LIRGs,   
we have started a program to  obtain IRS
spectral mapping of a sample of local LIRGs from Alonso-Herrero et
al. (2006). In this paper we present 
the first results for  Arp~299 
(IC~694 + NGC~3690), a LIRG  with 
$L_{\rm IR} \sim 6 \times 10^{11}\,{\rm L}_\odot$, 
using the {\it IRAS} fluxes from  
Sanders et al. (2003) and a distance of $d=42\,$Mpc ($H_0=75\,{\rm km
  \,s}^{-1}\,{\rm Mpc}^{-1}$). 

Arp~299 has long been  recognized as an extremely interesting system: 
an interacting galaxy (Arp 1966);  a non-Seyfert\footnote{but see
Garc\'{\i}a-Mar\'{\i}n et al. 2006 for the optical identification of the AGN
in NGC~3690}  galaxy
with bright   emission lines (Weedman 1972); a luminous 
mid-IR source   (Rieke \& Low 1972);  and a highly powerful starburst
 (Gehrz, Sramek, \& Weedman 1983).  The bright
sources in this system probe a full range of distinct and well-characterized 
physical conditions that can be used as a benchmark for different mid-IR
indicators for LIRGs. In brief, the properties of the 
nuclear region of IC~694, referred to
as A (see Gehrz et al. 1983), are dominated by a relatively
extended and evolved period of star 
formation, whereas the C+C$^\prime$ complex in the
overlap region between the two galaxies, contains powerful and young
starbursts. The nuclear region of NGC~3690, referred to as B1, hosts an obscured
AGN and is also surrounded by regions of star formation. 
\S5, 6, and 7 discuss 
the properties of these sources. Prior to that discussion, 
\S2 presents the
observations and  data reduction, whereas \S3 presents the analysis of the
data. In \S4 we describe the overall morphological mid-IR properties of the
system. Finally, \S8 puts the mid-IR properties of Arp~299 in
the context of local and high-$z$ IR-bright galaxies, and \S9 gives the
conclusions.

\section{Observations}

\subsection{IRS Spectral Mapping Observations}

The observations of the interacting system Arp~299 were obtained 
with IRS 
using all four modules: Short-High (SH; $9.9-19.6\,\mu$m), 
Long-High (LH; $18.7-37.2\,\mu$m), Short-Low
(SL1; $7.4-14.5\,\mu$m  and SL2; $5.2-7.7\,\mu$m) 
and Long-Low (LL1; $19.5-38\,\mu$m and LL2; $14.0-21.3\,\mu$m). 
The low resolution modules contain two subslits, one for the first
order (SL1 and LL1) and the other for the second 
order (SL2 and LL2). The high resolution modules
are cross dispersed so that ten orders fall on the array (see Houck et 
al. 2004 for full details). 
The IRS high ($R\sim 600$) spectral resolution data (SH and LH 
 modules) were obtained in Cycle 3 as part of a larger GTO program (P.I.:
 G. H. Rieke) aimed at  
obtaining {\it Spitzer}/IRS spectral 
mapping of a representative number of local LIRGs from the volume-limited
sample described by Alonso-Herrero et al. (2006). 
The spectral mapping data in low ($R\sim 60-126$)
spectral resolution (SL and LL modules)  
were taken from the {\it Spitzer}
archive and were part of a different GTO program (P.I.: J. R. 
Houck) observed in  Cycle 1.

\begin{figure*}

\includegraphics[width=15.cm,angle=-90]{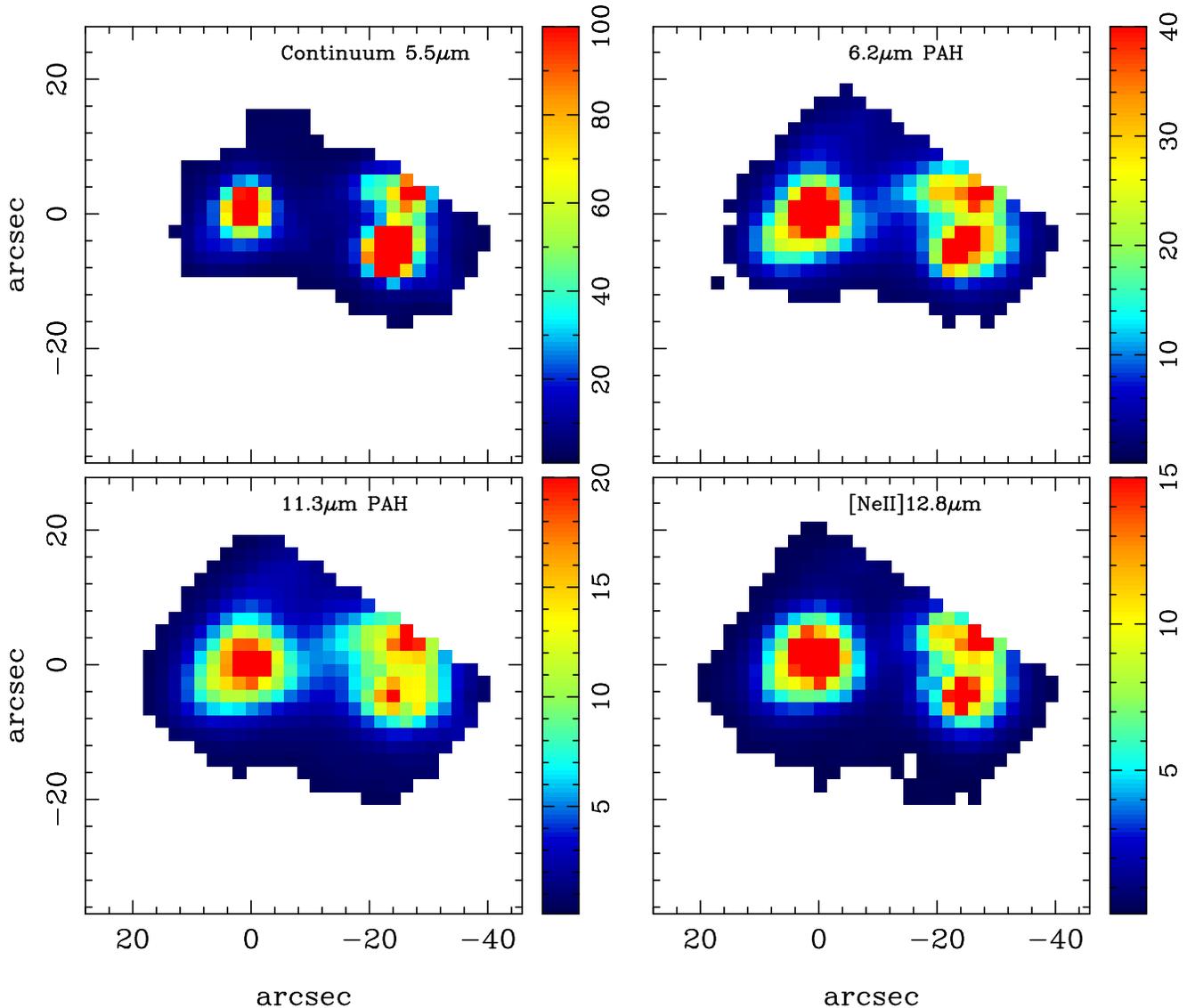}

\caption{Observed (not corrected for extinction) SL module spectral 
maps built with {\sc cubism} for the 
$5.5\,\mu$m continuum, 
the  $6.2\,\mu$m PAH feature, the 
$11.3\,\mu$m PAH complex ($11.2 + 11.3\,\mu$m PAH
features),  and 
the [Ne\,{\sc ii}]$12.81\,\mu$m emission line. For the PAHs and the 
emission line
the nearby continuum was fitted and subtracted. 
The units of the maps are MJy sr$^{-1}$ for the $5.5\,\mu$m continuum 
image, and 
$10^{-4}\,{\rm erg \, cm}^{-2}\,{\rm s}^{-1}\,{\rm sr}^{-1}$ for the
PAH features and the [Ne\,{\sc ii}] line maps. 
The orientation of the images
is north up, east to the left. In all the maps, the (0,0) point corresponds to
the approximate location of the nuclear region of IC~694. 
The images are shown in a linear scale. 
Only pixels with relative errors $\Delta f/f<0.2$ are displayed.}

\end{figure*}

\begin{figure*}

\hspace{1cm}
\includegraphics[width=8cm,angle=-90]{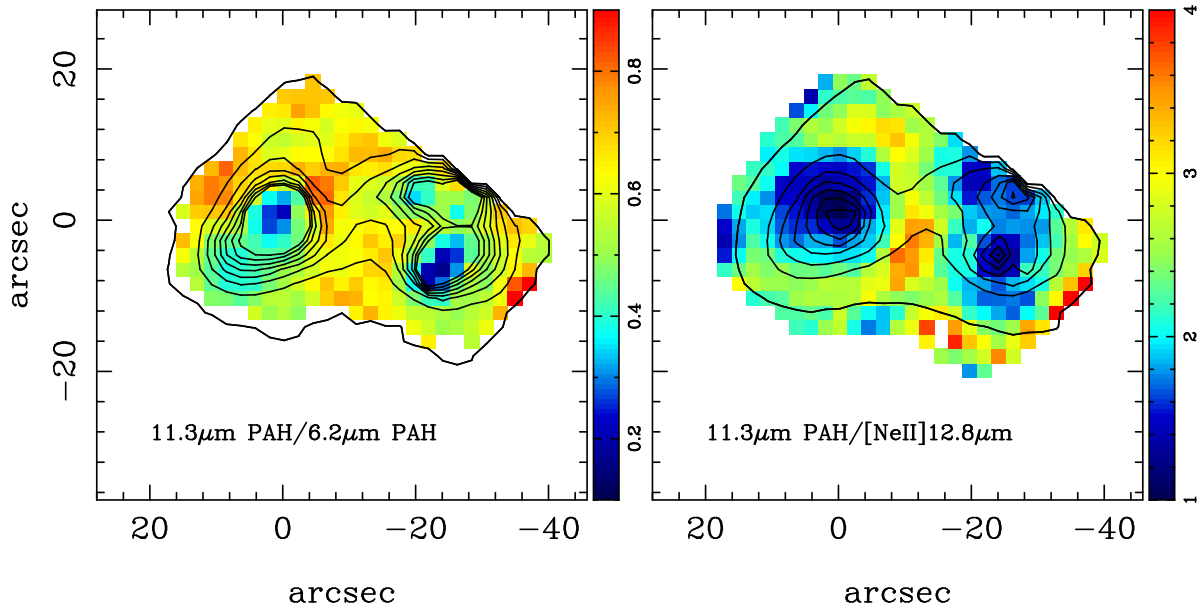}
\caption{Maps of the line ratios (not corrected for extinction)
  observed with the SL module. 
The orientation of the images
is north up, east to the left. The images are shown in a linear scale. The
  overlaid contours are the $6.2\,\mu$m PAH feature 
emission (left panel) and the 
$11.3\,\mu$m PAH complex emission (right panel).}
\end{figure*} 

The observing
strategy of both programs used the IRS spectral mapping
capability. In this observing mode 
the telescope is moved perpendicular to the long axis of the slit
with a step of one-half the slit width until 
the  appropriate region is covered. This strategy
produced a map redundancy of 2 for the SL, LL, and LH modules, and of 4 for
the SH module. The two galaxies of Arp~299
were observed as separate targets with the SH and LH modules, whereas in low
resolution they were observed with one pointing. Table~1 gives details of the
IRS spectral mapping 
observations, including the ramp integration times, number of cycles, and
final sizes of the data cubes. 

Since the SL and LL slits are longer than the
extent of the galaxy,  no separate 
background observations were needed. For the LH module we obtained dedicated
background observations with a single slit (staring mode) observation at a
region about 2 arcminutes away from the galaxy. No 
backgrounds were observed for the SH module. We found however, that even 
for the LH module where the background is higher than for the SH
module, the nuclei are so bright that the background subtraction had
little effect. 

The data were processed using the {\it Spitzer} IRS pipeline version S15.3.
The typical uncertainties of the wavelength calibration 
are $0.009\,\mu$m for SL1, 
$0.006\,\mu$m for SL2, $0.036\,\mu$m for LL1, $0.034\,\mu$m for LL2,
$0.003\,\mu$m for SH, and $0.01\,\mu$m for LH. The uncertainty of the absolute
photometric calibration is  less than $10\%$. The observations are diffraction
limited above $6\,\mu$m, so the spatial resolution depends on the wavelength
of the observation and it is generally $1.5-2.5$ pixels/spatial resolution
element (see IRS Handbook, version 3.1).  Our own estimates of the
angular resolution (FWHM) of the data cubes are typically $4\arcsec$ for SL2,
$5\arcsec$ for SL1, and  $5.5\arcsec$ for
SH (see Pereira-Santaella et al. 2009, in preparation). For the
assumed distance to Arp~299 these correspond to angular 
resolutios of 0.8\,kpc, 1\,kpc, and
1.1\,kpc for the SL2, SL1, and SH spectral maps, respectively. 

The data cubes were assembled using {\sc cubism}
(the CUbe Builder for IRS Spectra Maps, Smith et al. 2007b) from the
individual Basic Calibrated Data (BCD)  
spectral images. Full error cubes are also built alongside the data cubes 
by standard error propagation, using, for the input uncertainty, 
the BCD-level uncertainty estimates produced by the IRS pipeline  
from deviations of the fitted ramp slope fits for each pixel. These
uncertainties are used to provide error
estimates for extracted spectra, and constructed line and continuum maps (see
Smith et al. 2007b for full details). These errors together with 
the additional systematic
uncertainties arising from calibration effects and pointing errors
bring the error budget to typically $15-20\%$.

\subsection{IRS Staring Spectroscopy}

Additional IRS staring mode spectroscopy with the SH and LH modules of the 
two nuclei of Arp~299  was obtained from the {\it Spitzer} archive. These
observations were observed as part of 
Program 21, and had no associated background observations. 
We downloaded the BCD processed by
the {\it Spitzer} IRS pipeline 
(version S15.3 for SH and S17.2 for LH). After removing 
rogue pixels with IRSCLEAN, we extracted a spectrum from each BCD using 
SPICE. In doing so, SPICE
integrates the flux from the entire slit and performs an aperture
correction, which is dependent on the wavelength. We assumed point source flux
calibration for the extraction. The final step was to  
median combine all the spectra for each of the two nuclei and to correct for
fringing using IRSFRINGE.

\section{Analysis}

\subsection{Spectral Maps}
For the SL data cubes we used {\sc cubism} to construct spectral maps of the 
$6.2\,\mu$m aromatic (hereafter PAH) feature  
and the $11.3\,\mu$m PAH complex ($11.2 + 11.3\,\mu$m
PAHs), the [Ne\,{\sc ii}]$12.81\,\mu$m emission
line, as well as the continuum at $5.5\,\mu$m.  The integrated
line flux maps are computed by integrating  the line flux over 
the user-defined emission line regions with an average continuum removed.  
The $5.5\,\mu$m continuum map was constructed as the mean flux over a
bandpass covering the spectral range of $5.3-5.7\,\mu$m. The map 
of the $6.2\,\mu$m PAH feature was constructed by integrating the
feature in the $6.1-6.6\,\mu$m range and fitting the continuum 
between 5.3 and $6.9\,\mu$m, whereas the $11.3\,\mu$m PAH feature was 
integrated between 11.1 and $11.8\,\mu$m and the corresponding
continuum was fitted between $10.8$ and $12.1\,\mu$m. Finally the [Ne\,{\sc ii}]
emission line was integrated between $12.8$ and $13.1\,\mu$m, and 
the continuum was fitted between 12.2 and $13.4\,\mu$m. All these
wavelengths are observed values. 

Since {\sc cubism} does not fit or deblend  emission lines, the SL 
[Ne\,{\sc ii}]$12.81\,\mu$m map includes some contribution from the
$12.7\,\mu$m 
PAH feature complex. 
The SL spectral maps were then trimmed to the galaxy field of view (FOV) and 
rotated to the usual orientation of north  up, east to the left. 
As discussed in \S2.1  
{\sc cubism}  produces both  emission line and uncertainty maps.
These spectral maps, not corrected for extinction, are shown in Fig.~1. In
addition, we constructed maps of the observed 
$6.2\,\mu$m to $11.3\,\mu$m PAH ratio
and the $11.3\,\mu$m PAH to [Ne\,{\sc ii}]$12.81\,\mu$m ratio 
(see Fig.~2). Before constructing the line ratio
  maps the individual images were clipped to those regions with relative
  errors of $\Delta f/f < 0.2$ (or S/N$>2$), where $f$ and
  $\Delta f$ are the integrated line flux and associated uncertainty,
  respectively, as derived by {\sc cubism}. The typical S/N of the
  bright nuclei in the SL maps are between 400 and 700 per pixel.

\begin{figure*}
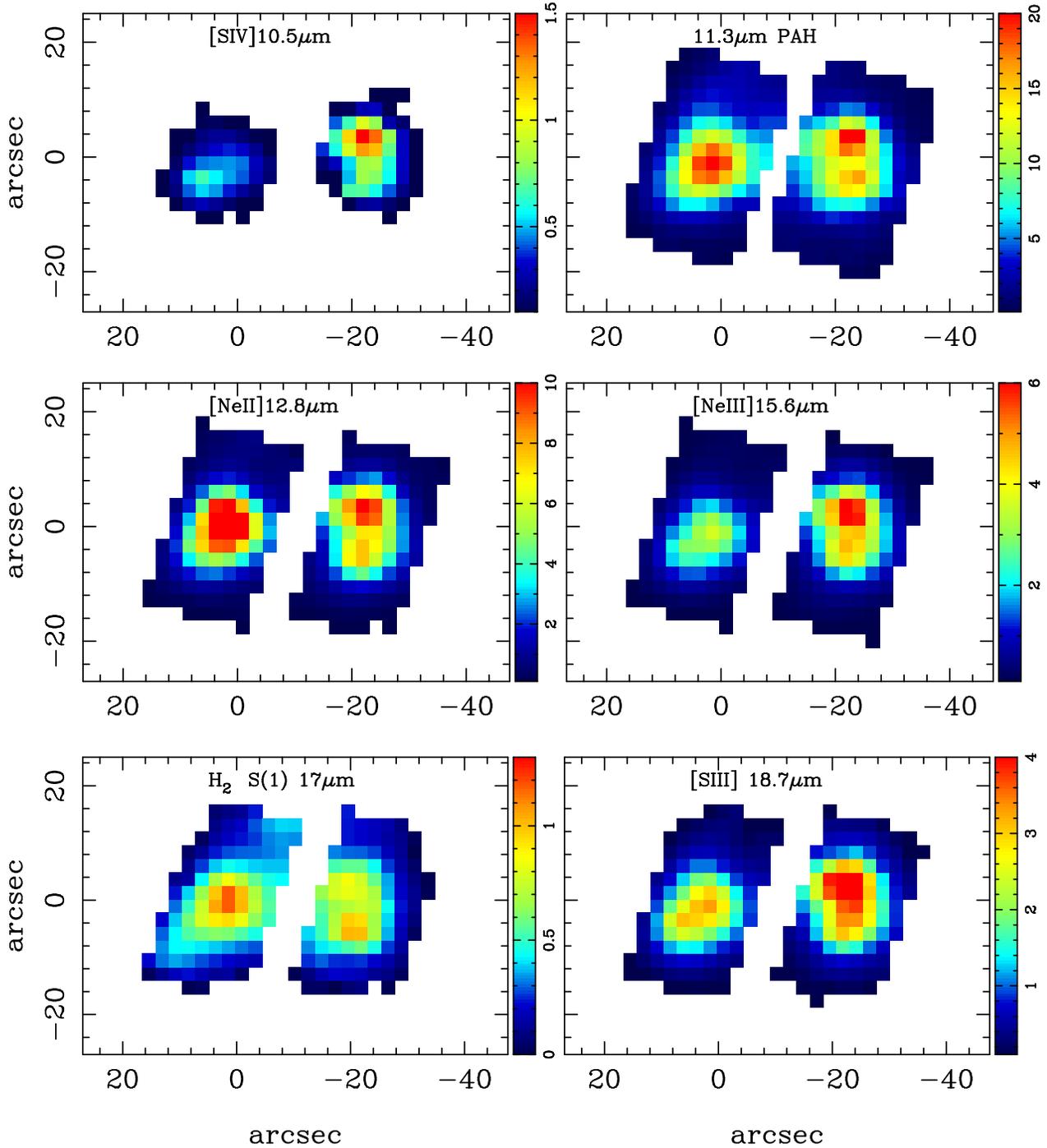


\includegraphics[width=11.6cm,angle=-90]{f3a.ps}

\vspace{0.5cm}

\includegraphics[width=6.4cm,angle=-90]{f3b.ps}

\caption{
Mosaic of the SH module observed (not corrected for extinction) 
 continuum-subtracted emission lines maps, as well as the 
$11.2 + 11.3\,\mu$m PAH complex. 
The individual maps for the two pointings (Table~1) 
were built by fitting simultaneously the nearby continuum and Gaussians to the
emission lines. Only pixels where the lines are detected at the $3\sigma$
level or higher are displayed.
The units of these maps are $10^{-4}\,{\rm erg \,cm}^{-2}\,{\rm
  s}^{-1}\,{\rm sr}^{-1}$.  
The spectral map of the $11.3\,\mu$m PAH feature was built
with {\sc cubism} as explained in \S3.1, and is shown in the 
same units as the other SH spectral maps. The orientation of all the images
is north up, east to the left. The images are shown in a linear scale. }

\end{figure*} 

\begin{figure*}

\hspace{1.cm}
\includegraphics[width=6cm,angle=-90]{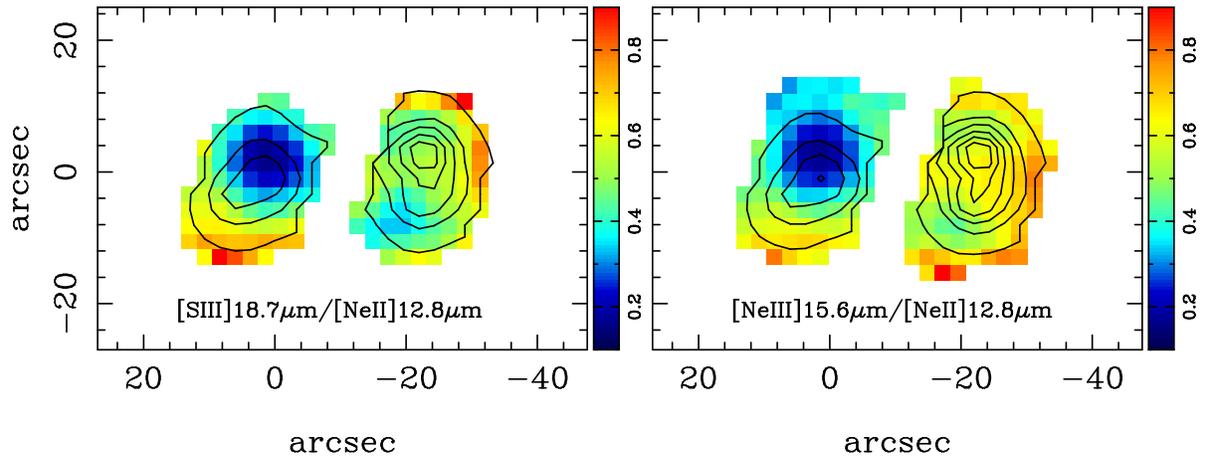}

\caption{Mosaics of line ratios (not corrected for extinction) observed with
  the SH module. 
The images are shown in a linear scale. The
  overlaid contours are the [S\,{\sc iii}]$18.71\,\mu$m line emission (left
  panel) and the [Ne\,{\sc iii}]$15.56\,\mu$m line emission (right panel).
}

\end{figure*}

\begin{figure*}
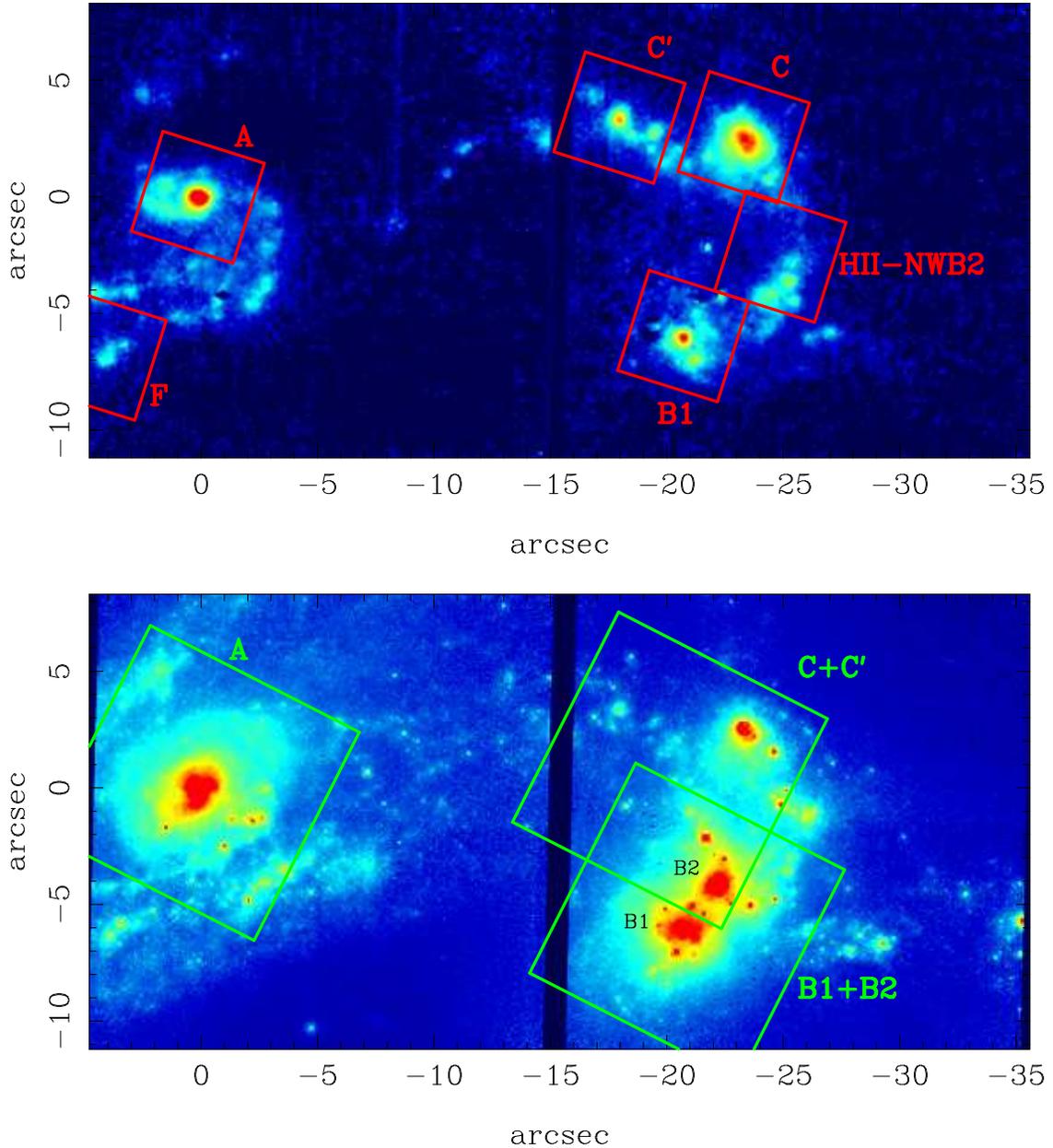


\hspace{1.5cm}
\includegraphics[width=8cm,angle=-90]{f5a.ps}

\vspace{0.5cm}

\hspace{1.5cm}
\includegraphics[width=8cm,angle=-90]{f5b.ps}

\caption{{\it HST}/NICMOS images of the continuum-subtracted
  Pa$\alpha$ emission at
  $1.875\,\mu$m  
(top panel) and of the continuum at $1.6\,\mu$m (bottom panel)  of Arp~299
  (see AAH00). The small squares (upper panel) and large squares
  (bottom panel) are the
  approximate locations, orientations, and sizes  
of the extraction apertures for the SH ($\sim
  4.5\arcsec \times 4.5\arcsec$) and the
  low-resolution (LL and matched SL apertures, $\sim 10.2\arcsec \times
  10.2\arcsec$) spectra, respectively.  
The orientation of the images
is north up, east to the left. The images are shown in a square root
scale. We also mark the positions of the two bright $1.6\,\mu$m
continuum sources B1 and B2 in the lower panel. 
Note that there is very little, if any, Pa$\alpha$ emission 
arising from B2 (upper panel). }
\end{figure*}

For the bright emission lines present in the SH data cubes 
([S\,{\sc iv}]$10.51\,\mu$m, [Ne\,{\sc
      ii}]$12.81\,\mu$m, [Ne\,{\sc iii}]$15.56\,\mu$m, [S\,{\sc
      iii}]$18.71\,\mu$m, and the H$_2$ S(1) line at $17.04\,\mu$m), 
we developed our own {\sc idl} routines to produce spectral maps. 
Briefly, we first automatically extracted 
1D spectra in $2 {\rm pixel} \times {\rm 2 pixel}$ boxes, by moving
pixel by pixel along columns and rows until the entire FOV was covered. Each 
emission line was fitted with a 
Gaussian, and the continuum was fitted from adjacent spectral regions. In the
case of the [Ne\,{\sc ii}]$12.81\,\mu$m line 
we used two Gaussians, the second one 
to account for the nearby $12.7\,\mu$m PAH feature. The lines were
only fitted when the peak of the emission line
was $>3\sigma$ above the local continuum, where 
$\sigma$ is the standard deviation
of the local continuum. If this requirement was not met, the pixel or pixels
were masked out in the final spectral map. Since the $11.3\,\mu$m PAH feature
is not well reproduced with a Gaussian function, we used {\sc cubism} instead,
as described for the SL data cubes, to construct the spectral maps. 
In this case, the $11.3\,\mu$m PAH
feature was integrated between 11.2 and $11.7\,\mu$m, and the continuum
was fitted between 10.7 and $12.3\,\mu$m.

After rotation of the SH spectral
maps, we constructed  mosaics with the two pointings (Table~1) obtained for
the system as shown in Fig.~3. We also constructed maps of the 
[S\,{\sc iii}]$18.71\,\mu$m/[Ne\,{\sc ii}]$12.81\,\mu$m and [Ne\,{\sc
      iii}]$15.56\,\mu$m/[Ne\,{\sc
      ii}]$12.81\,\mu$m line ratios. We imposed a further restriction to
include a given pixel in the map that the line errors of each emission line
were less than 25\%. The maps of the two line ratios are shown in Fig.~4.

\begin{figure}
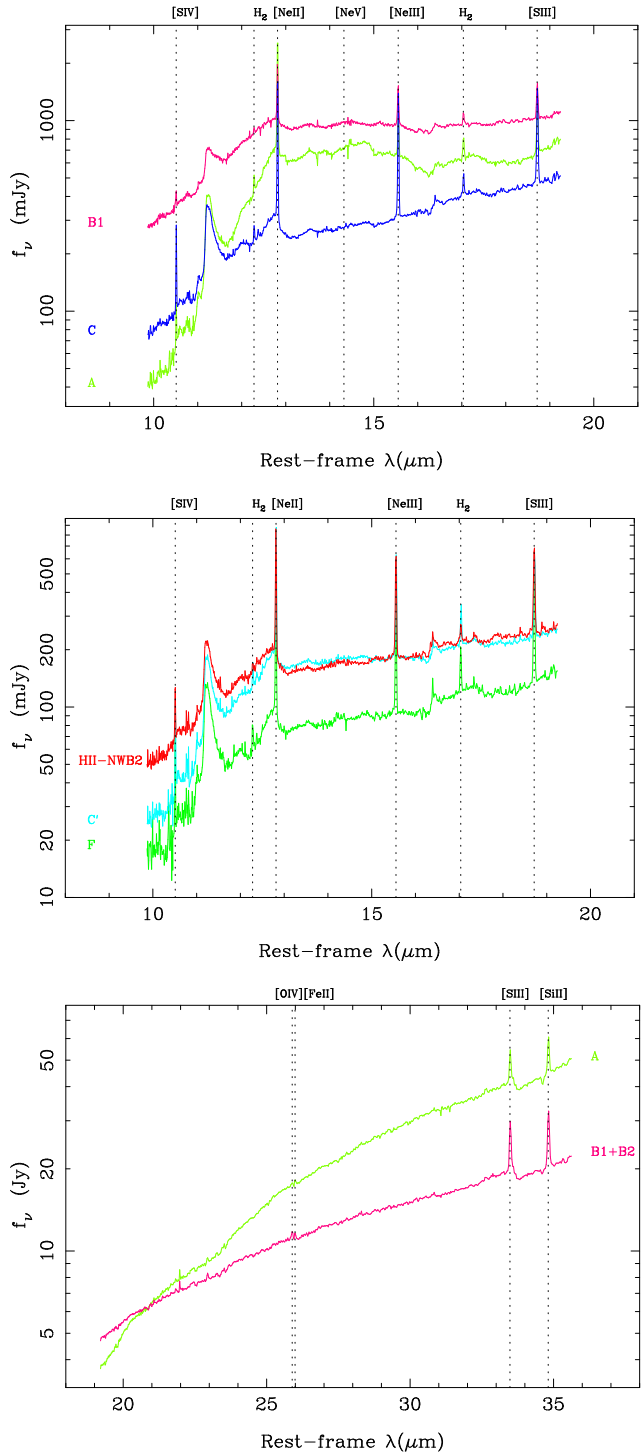


\includegraphics[width=6.2cm,angle=-90]{f6a.ps}

\hspace{3cm}

\includegraphics[width=6.2cm,angle=-90]{f6b.ps}

\hspace{3cm}

\includegraphics[width=6.2cm,angle=-90]{f6c.ps}

\caption{IRS high spectral resolution 1D spectra     for selected sources in
  Arp~299. {\it Top and Middle:} 
SH spectra of six bright sources in Arp~299 extracted from the spectral
  mapping data cubes with 
apertures of $\sim 4.5\arcsec \times 4.5\arcsec$ (see Fig.~5). 
{\it Bottom:} Staring mode LH spectra of the nuclear regions of IC~694 and 
NGC~3690 extracted as point sources.  
The locations of the brightest
emission lines are marked.}

\end{figure} 

\subsection{Extraction of the 1D Spectra}

For the two short wavelength 
IRS modules we first used {\sc cubism} with the smallest ($2
\times 2$ pixel) possible  extraction apertures 
on the data cubes before rotation.
We extracted 1D SH and SL 
spectra for the brightest sources of Arp~299, namely A, B1, C, and  
C$^\prime$ (see Fig.~5). The optical 
nucleus B2 is known to be faint in the mid-IR (Keto et al. 1997; 
Soifer et al. 2001), so we
expect its contribution to the mid-IR spectrum of B1 to be minimal.
However, there are H\,{\sc ii} regions around B1 (see Fig.~5) 
that  contribute to the SH and SL mid-IR spectra. 

Additionally we extracted SH spectra for two bright H\,{\sc ii}
complexes. The first  is in the  
 southern spiral arm of IC~694, about 7\arcsec \ southeast of the 
nucleus, and it is clearly seen 
in the [Ne\,{\sc iii}]$15.56\,\mu$m, 
[S\,{\sc iii}]$18.71\,\mu$m, and [S\,{\sc iv}]$10.51\,\mu$m emission
line maps in Fig.~3. 
The location of this region corresponds approximately to 
source F identified in H$\alpha$ by Garc\'{\i}a-Mar\'{\i}n et al. (2006) as one
of the youngest H\,{\sc ii} regions in IC~694, based on the equivalent width 
(EW) of
H$\alpha$. This region  was also detected
in the  radio H92$\alpha$ emission line by Zhao  et al. (1997). 
The second  complex is located in NGC~3690, about 
3\arcsec \ northwest from B2. 
In Fig.~5 the approximate locations of the SH 
extraction apertures are shown superimposed
on the {\it HST}/NICMOS continuum-subtracted Pa$\alpha$ image of
Alonso-Herrero et al. (2000, AAH00 hereafter). 
The SH spectra for a $4.5\arcsec \times 4.5\arcsec$ aperture 
of all these sources are shown in Fig.~6 (top and middle panels). 
The LH spectra of the three brightest regions A, B1+B2, and C+C$^\prime$
were extracted with a $\sim 13.4\arcsec \times 13.4\arcsec$ 
aperture. We used {\sc cubism} to extract 
matching SH spectra so that the full 
$\sim 9.9-37\,\mu$m spectral range was covered. 
The staring mode LH spectra of A and B1+B2 are shown in the lower panel of
Fig.~6. 

Similarly, LL spectra  with $\sim 10.4\arcsec \times 10.4\arcsec$
apertures  
were obtained for the brightest sources of Arp~299 (see Fig.~5),
together with 
matching SL spectra so  full $\sim 5-38\,\mu$m low-resolution spectra
of A, B1+B2, and C+C$^\prime$ were obtained, although there is some
overlap between the B1+B2, and the C+C$^\prime$ extraction apertures. 
Finally, we extracted a 1D
low-resolution (SL+LL) spectra 
covering approximately the whole system  using  an LL aperture (before
rotation) of 
$ 6 \times 8\,$pixel ($\sim 30.5\arcsec \times 40.6\arcsec$). All the
SL+LL spectra are shown in  Fig.~7.

\subsection{Spectral features in the low-resolution spectra}
The low-resolution spectra (Fig.~7) for the three main components of the
system (A, B1+B2, and C+C$^\prime$) 
as well as the integrated spectrum were fitted
with {\sc pahfit} (Smith et al. 2007a). {\sc pahfit} uses a simple,
physically-motivated model that incorporates 
starlight, thermal dust continuum in
a small number of fixed temperature bins, resolved dust features and feature
blends, and prominent emission lines, plus  fully-mixed 
or screen dust extinction, including 
the silicate absorption bands at 9.7 and $18\,\mu$m.  {\sc pahfit}
is designed to work with the low spectral resolution data, and
provides the best results when the full $5-40\,\mu$m spectral range is
used. 

In Table~2 we give the {\sc pahfit} results for the brightest emission line
ratios (corrected for extinction), for the EW of the
$6.2\,\mu$m PAH feature, and for the 
optical depth of the $9.7\,\mu$m silicate feature from the two dust
configurations. The mid-IR extinction law used by {\sc pahfit} can be found in
Smith et al. (20071), The optical depth can be
converted to optical 
extinction ($A_V$) using the conversion $A_V/\tau_{\rm Si}=16.6$
from Rieke \& Lebofsky (1985). The two dust geometries provide 
fits of similar quality
(see Table~2), with the screen model having lower silicate depths  
(between 2 and 3) than the fully-mixed model. This behavior is similar to that
found by Smith et al. (2007a) for the SINGS galaxies. Unlike the majority of
the SINGS galaxies, the Arp~299 
spectra are fitted with high values of extinction, going from $A_V\sim
12$ to $A_V\sim 33\,$mag
(screen model geometry) for the different components of the system
(see Table~2).

\begin{figure}
\includegraphics[width=6.5cm,angle=-90]{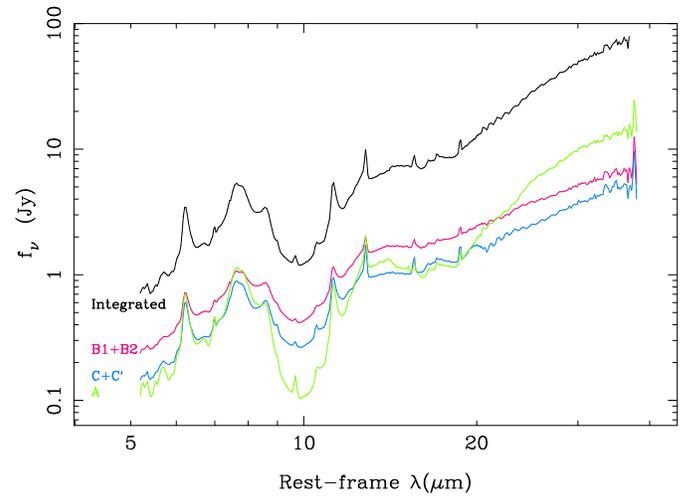}

\caption{Low spectral-resolution (SL+LL) spectra of 
the two nuclei of Arp~299 (A and B1+B2), the 
C+C$^\prime$  complex,  
as well as the integrated spectrum for the interacting
system. The extraction apertures (shown in Fig.~5) are $10.4\arcsec \times
10.4\arcsec$ for A, B1+B2, and C+C$^\prime$, and 
$30.5\arcsec \times 40.6\arcsec$ for the integrated spectrum. See
\S3.2 for details.} 

\end{figure}

\subsection{Spectral features in the SH and LH spectra}
The fluxes of the brightest emission lines in the SH and LH spectra were measured
using {\it splot} within the {\sc iraf} {\it noao} environment. For all the
emission lines the continuum was 
estimated from regions adjacent to the line. Most of the lines are well
separated in the SH module, except for the [Ne\,{\sc ii}]$12.81\,\mu$m line. In
this case, to get a good estimate of the line fluxes and the local continuum, 
we fitted simultaneously three Gaussians to the H$_2$ S(2) line at
$12.28\,\mu$m, the $12.7\,\mu$m PAH complex and the [Ne\,{\sc ii}] line. 

The observed (not corrected for extinction) small-aperture 
SH  line ratios in the six bright components 
described in \S3.2 are listed in Table~3. We also give in this table 
the [Ne\,{\sc ii}]$12.81\,\mu$m line fluxes and errors, 
because it is the line with the
largest errors as it is blended with the broad $12.7\,\mu$m PAH
feature. Additional uncertainties in the fluxes and line ratios are due
to centering errors and  possible aperture corrections. As 
an example of the latter issue, in  
the notes to this table we also give the [Ne\,{\sc ii}] and
[Ne\,{\sc iii}] fluxes for the nuclear region A of IC~694 measured
with the SH module through a
slightly larger extraction aperture ($3 {\rm pixel} \times 3 {\rm
  pixel} \sim 6.8\arcsec \times 6.8\arcsec$). The fluxes are about a
factor of 1.6 larger indicating that some flux is missing from the $4.5\arcsec
\times 4.5\arcsec$ aperture, 
although the Pa$\alpha$ emission (Fig.~5), radio continuum and line
nuclear emissions (Aalto et al. 1997; Zhao et al. 1997) 
appear to be coming from a region  less than 
5\arcsec \, in diameter.

The observed line ratios and [Ne\,{\sc ii}] line fluxes measured
from the SH+LH spectra through $13.4\arcsec \times 13.4\arcsec$
apertures are given in Table~4 for the three main components of the
system. The fluxes and line ratios for IC~694-A, and the sum of
 NGC-3690-B1+B2 and 
 NGC-3690-C+C$^\prime$ are in good agreement with those reported by Verma et
 al. (2003) from {\it ISO} large-aperture 
spectroscopy. We also note that the line
 ratios given in Table~4 are consistent within the uncertainties with
 those from the SL+LL data (Table~2) if they are corrected for
 extinction as derived from {\sc pahfit} and the dust screen 
model.

\section{Overall morphological properties}
In this section we describe the overall morphology of
the different mid-IR spectral features, as well as the general physical and
excitation  conditions in Arp~299.

\subsection{Mid-IR continuum and fine-structure lines}
 Figures~1 and 3 show that there is bright mid-IR continuum and  
line emission from the two nuclei of 
the system (A and B1), as well as from the C+C$^\prime$ complex in the overlap
region between the two galaxies. There is also emission in the 
spiral arms of IC~694 (especially in
the southeast spiral arm), as well as in regions west of B1 and B2 in
NGC~3690. These results are in agreement with previous work using
optical, near-IR, mid-IR, millimeter, 
and radio continuum and 
emission lines (e.g., Gehrz et al. 1983; Aalto et al. 1997; 
Zhao et al. 1997; 
Satyapal et al. 1999; Sugai et al. 1999; AAH00; 
Soifer et al. 2001; 
Gallais et al. 2004; Garc\'{\i}a-Mar\'{\i}n et al. 2006; Imanishi \&
Nakanishi 2006). 

In particular, the morphology of the fine-structure 
mid-IR  [Ne\,{\sc iii}]$15.56\,\mu$m and 
[S\,{\sc iii}]$18.71\,\mu$m emission lines as well as the relative
brightnesses from source to source 
resemble  that of the {\it HST}/NICMOS 
Pa$\alpha$ emission line (Fig.~5, and AAH00) and
other near-IR hydrogen recombination lines (Sugai et al. 1999; Satyapal et
al. 1999). The hydrogen recombination 
lines are tracers of the young 
ionizing stellar populations. The [S\,{\sc iv}]$10.51\,\mu$m line
shows an overall morphology that is similar to 
Pa$\alpha$, but it is more affected by extinction,
especially toward the nuclear region of IC~694 as we shall see in
\S7.1.

\begin{table*}
\caption{Results from {\sc pahfit} for SL+LL spectra}

\small

\begin{tabular}{lcccccccccc}
\hline
Source & $\tau_{\rm Si}$ 
& $\frac{ {\rm [NeIII]}15.6\mu{\rm m}}{{\rm [NeII]}12.8\mu{\rm m}}$  
& $\frac{ {\rm [ArIII]}9.0\mu{\rm m}}{{\rm [ArII]}7.0\mu{\rm m}}$  
& $\frac {{\rm [SIV]}10.5\mu{\rm m}} {{\rm [SIII]}18.7\mu{\rm m}}$
& $\frac {{\rm [SIII]}33.5\mu{\rm m}} {{\rm [SIII]}18.7\mu{\rm m}}$
& $\frac{ {\rm [SIV]}10.5\mu{\rm m}}{{\rm [ArIII]}9.0\mu{\rm m}}$  
& $\frac{ {\rm [SIII]}18.7\mu{\rm m}}{{\rm [NeII]}12.8\mu{\rm m}}$  
& EW 
& $\frac{ {\rm PAH \,}6.2\mu{\rm m}}{{\rm PAH}11.3\mu{\rm m}}$  
& $\chi^2$\\
&&&&&&&&$6.2\,\mu$m  PAH
\\
\hline
\multicolumn{11}{c}{Screen Model}\\
 \hline
A          & 2.04 & 0.06 & 1.46 & 0.47 & 0.54 & 0.37 & 0.27 & 1.36 & 1.5 & 
1535\\
B1+B2      & 0.91 & 0.32 & 3.00 & 0.03 & 0.99 & 0.04 & 0.57 & 0.22 & 1.8 & 
1012 \\
C+C$^\prime$   & 0.78 & 0.34 & 1.56 & 0.32 & 1.16 & 0.46 & 0.40 & 0.50 & 1.6 & 
670\\
Integrated & 1.09 & 0.36 & 1.57 & 0.15 & 1.35 & 0.38 & 0.48 & 0.65 & 1.5 & 
1418\\
\hline
\multicolumn{10}{c}{Mixed Model}\\
 \hline
A          & 8.75 & 0.05 & 1.65 & 0.35 & 0.25 & 0.33 & 0.39 & 1.85 & 1.2 & 1139 \\
B1+B2      & 2.42 & 0.32 & 3.68 & 0.06 & 0.86 & 0.12 & 0.62 & 0.22 & 1.6    & 823 \\
C+C$^\prime$    & 1.98 & 0.34 & 1.68 & 0.35 & 1.07 & 0.49 & 0.43 & 0.50 & 1.5 & 627\\
Integrated & 2.98 & 0.34 & 1.84 & 0.20 & 1.23 & 0.45 & 0.54 & 0.65 & 1.4   & 1170\\
\hline
\end{tabular}

\tablecomments{
The extraction apertures are $10.4\arcsec \times
10.4\arcsec$ for A, B1+B2, and C+C$^\prime$, and 
$30.5\arcsec \times 40.6\arcsec$ for the integrated spectrum. 
The line ratios are corrected for extinction as derived from the
measured $9.7\,\mu$m optical depth and using the extinction law 
of Smith et al. (2007a).  The EW of the $6.2\,\mu$m PAH
feature, defined as 
EW$=f(6.2\,\mu{\rm m \,PAH})/f(6.2\,\mu{\rm m \,continuum})$, 
are measured in $\mu$m. The value of $\chi^2$ given in the last 
column indicates the goodness of the fits to the spectra done 
with {\sc pahfit}.}

\end{table*}

\subsection{The PAH Emission}

The PAH emission shows some morphological differences when compared to the
mid-IR  fine structure lines, as can be seen from Figs.~1 and 3.  
In addition to the aforementioned bright sources in this system, 
there is diffuse bright 
PAH emission at 6.2 and $11.3\,\mu$m (Fig.~1) in
the interface region connecting the two galaxies, southeast of C+C$^\prime$. 
There is relatively bright PAH diffuse emission in the interface region where
there is very little Pa$\alpha$ emission (Fig.~5), possibly indicating that
some PAH emission is not associated with the 
ionizing young
stellar populations. 

Theoretical models predict that the relative strengths of the different PAH
features depend on the properties of the dust grains, the charging conditions,
as well as the starlight 
intensity (Draine \& Li 2001). In particular, neutral PAHs are
expected to show large 
$11.3\,\mu$m to $6.2\,\mu$m ratios, whereas ionized PAHs have smaller ratios. 
The map showing the $11.3\,\mu$m to $6.2\,\mu$m PAH ratio (Fig.~2, left
panel), shows a range\footnote{The method used to construct the SL spectral
  maps for the PAH features is equivalent to a spline fitting, as
  opposed to full fitting of the spectra (see \S3.3). In general Smith et
  al. (2007a) find that for the 6.2 and $11.3\,\mu$m PAH features the full
  decomposition method measures intensities of about a factor of two higher
  than the spline method for both these two PAH features. We can then 
assume that
  the  $11.3\,\mu$m to $6.2\,\mu$m PAH ratio map should give comparable
  results to the ratios fitted with {\sc pahfit}.} from approximately 
0.2 to 0.9.
From this figure it appears, as predicted by the models, that the lowest PAH
ratios are associated with the regions of high ionization (i.e., 
high [Ne\,{\sc iii}]/[Ne\,{\sc ii}] line ratios) in the galaxy. In
comparison, the
regions with high $11.3\,\mu$m to $6.2\,\mu$m PAH ratios are to the northwest
of A and to the east of B1 where there is no evidence for bright Pa$\alpha$
emission associated with 
ionizing stellar
populations (Fig.~5). We note however, that both this ratio 
and the $11.3\,\mu$m PAH/[Ne\,{\sc ii}]$12.81\,\mu$m ratio 
are subject to extinction effects. This is because the $9.7\,\mu$m
absorption feature is broad, and the $11.3\,\mu$m feature is inside it 
(see figure~3 of Smith et al. 2007a for the mid-IR extinction law of
dust similar to that of the Milky Way). 
Thus, caution should be exercised when interpreting ratios involving 
the $11.3\,\mu$m PAH feature
(see discussions by Beir\~ao et al. 2008
and Pereira-Santaella et al. 2009). For example, 
the observed values for the nuclear regions of
both galaxies (A and B1) are probably lower limits as both nuclei show
significant values of extinction (\S5 and \S7.1). 

\subsection{Molecular Gas}

The $17\,\mu$m H$_2$ S(1) emission from IC~694 shows a more disk-like
morphology 
compared to the mid-IR fine-structure 
emission lines, which mainly trace H\,{\sc ii}
emission in the spiral arms. The disk scenario for IC~694 is further
confirmed by the observed H$_2$ velocity field (Pereira-Santaella et
al. 2009). NGC~3690 shows H$_2$ emission from the B1+B2
area, and the C+C$^\prime$ complex. The 
H$_2$ S(1) emission of the whole system
follows very closely the morphology of the molecular gas as traced by the 
$^{12}$CO $1-0$ line (Aalto et al. 1997; Casali et al. 1999), 
as also found by Sugai et al. (1999)
for the near-IR H$_2$ transitions.

\subsection{Physical and Excitation Conditions of the gas}

The [S\,{\sc iii}]$18.71\,\mu$m/[Ne\,{\sc ii}]$12.81\,\mu$m line ratio
is almost insensitive to the hardness of the radiation field (i.e., age of the
stellar population) and decreases with increasing densities,
especially in  the  $n_{\rm H} \sim 10^3-
10^6\,{\rm cm}^{-3}$ range, depending on the ionization parameter
(see Snijders et al. 2007). The SH
observed [S\,{\sc iii}]$18.71\,\mu$m/[Ne\,{\sc ii}]$12.81\,\mu$m line
ratio map (Fig.~4) 
shows spatial variations on physical scales of 
$\sim 1\,$kpc (${\rm FWHM} \sim 2\,$pixels). Leaving
aside the nuclear region of IC~694\footnote{the extinction corrections to 
the [S\,{\sc iii}]$18.7\,\mu$m line (see extinction curve in Smith et
al. 2007a and table~2 in Farrah et al. 2007) are
non-negligible for this source (Table~2, and see also  \S7.1)}, 
the observed ratio varies between approximately 0.2 and 0.7. 

Perhaps not
surprisingly, those regions with the highest  
[S\,{\sc iii}]$18.71\,\mu$m/[Ne\,{\sc ii}]$12.81\,\mu$m ratios  do not
coincide with the brightest mid-IR emitting regions. Rather they 
are located mostly in
the southern part of IC~694, and to the north of the C+C$^\prime$ 
complex and south of
B1.  If this ratio traces the density (assuming a constant ionization
parameter, cf \S5 and 7) these would be the least dense regions in the system.
The nuclei of the galaxies (B1 and A) 
present the lowest ratios (see also Tables~3, 4, and 5), and in
particular A, the nuclear region of IC~694, would be the densest region,  
of the order of $n_{\rm H} \sim
10^3- 10^4\,{\rm cm}^{-3}$ (for solar metallicity, see Snijders et al. 2007). 
The significant change of the gas properties between the spiral arms 
and the nuclear region of IC~694 observed in the 
[S\,{\sc iii}]$18.71\,\mu$m/[Ne\,{\sc ii}]$12.81\,\mu$m spectral map is
consistent with the variations of the molecular line ratios (Aalto et
al. 1997). The molecular properties of A require 
a population of unusually dense and warm clouds (see
also Zhao et al. 1997).  The C+C$^\prime$ star-forming complex in the overlap
region of the two galaxies presents  
[S\,{\sc iii}]$18.71\,\mu$m/[Ne\,{\sc ii}]$12.81\,\mu$m line ratios
intermediate between those of the
dense nuclear regions and the extranuclear regions of low surface brightness. 


The  [Ne\,{\sc iii}]$15.56\,\mu$m/[Ne\,{\sc ii}]$12.81\,\mu$m line ratio 
is sensitive to the hardness of the radiation field and has been
quantitatively investigated by a number of works (e.g., Thornley et
al. 2000; Mart\'{\i}n-Hern\'andez et al. 2002; 
Verma et al. 2003; Rigby \& Rieke
2004; Snijders et al. 2007). 
 This ratio has advantages over
other mid-IR hardness-dependent line ratios, in that 
it is less affected by differential extinction than  
[S\,{\sc iv}]$10.5\,\mu$m/[S\,{\sc iii}]$18.71\,\mu$m, and  
can be measured from the SH spectra, unlike  
for instance [Ar\,{\sc iii}]$9.0\,\mu$m/[Ar\,{\sc ii}]$7.0\,\mu$m.

Figure~4 shows the SH map of the observed
[Ne\,{\sc iii}]$15.56\,\mu$m/[Ne\,{\sc ii}]$12.81\,\mu$m line ratio,
ranging from 0.2 to approximately
0.9. The two nuclei show the smallest ratios, 
whereas the  C+C$^\prime$ complex and the southern spiral arm of
IC~694, which are 
known to host some of the  youngest regions in the system, show 
intermediate values. The highest ratios, however appear to be
connected with regions of low surface brightness (e.g., the 
region about 10\arcsec
\ northwest of B1). It is not clear if the ratios in these regions 
reflect the youth of the stellar population, as they do not
seem to be associated with bright H\,{\sc ii} regions detected in 
Pa$\alpha$ (see Fig.~5), or
rather result from other processes. For instance, there is bright
diffuse UV emission to the west of sources B1 and B2 (Meurer et al. 1995). 
Moreover, a similar situation was observed in M82 where the 
[Ne\,{\sc iii}]$15.56\,\mu$m/[Ne\,{\sc ii}]$12.81\,\mu$m line ratio
increases at increasing distances form the galactic plane (Beir\~ao et
al. 2008). 

\begin{table*}
\caption{Observed Line Ratios from the SH spectral mapping data extracted with a $4.5\arcsec
  \times 4.5\arcsec$ aperture. }

\small

\begin{tabular}{lcccccccc}
\hline
Source & $f({\rm [NeII]}12.8\mu{\rm m})$ &
$\frac{ {\rm [NeIII]}15.6\mu{\rm m}}{{\rm [NeII]}12.8\mu{\rm m}}$ 
&  $\frac {{\rm [SIV]}10.5\mu{\rm m}} {{\rm [SIII]}18.7\mu{\rm m}}$  
&  $\frac {{\rm [SIII]}18.7\mu{\rm m}} {{\rm [NeII]}12.8\mu{\rm m}}$ 
& $\frac {{\rm H_2 S(2)}} {{\rm H_2 S(1)}}$ & 
$\frac {{\rm H_2 S(1)}}{{\rm [NeII]}12.8\mu{\rm m}}$ \\
\\
\hline
IC~694-A           &$11.0\pm 2.2$ & 0.15 & 0.13 & 0.16 & 0.65 & 0.07\\
IC~694-F           &$2.0\pm 0.4$ & 0.55 & 0.22 & 0.56 & 0.42 & 0.14\\
NGC~3690-B1(+B2)   &$4.6\pm 0.9$ & 0.44 & 0.25 & 0.33 & --- & 0.11\\
NGC~3690-C         &$6.0\pm 0.6$ & 0.60 & 0.28 & 0.49 & 0.45  & 0.07\\
NGC~3690-C$^\prime$ &$3.0\pm 0.3$ & 0.48 & 0.30 & 0.40 & 0.71 & 0.12 \\
NGC~3690-HII-NWB2  &$2.8\pm 0.6$ & 0.64 & 0.21 & 0.55 & --- & ---\\
\hline
\end{tabular}

\tablecomments{
The [Ne\,{\sc ii}] fluxes are in units of 
$10^{-13}\,{\rm erg \,cm}^{-2}{\rm s}^{-1}$. For comparison the
$6.8\arcsec \times 6.8\arcsec$ 
[Ne\,{\sc ii}] and [Ne\,{\sc iii}] fluxes for IC~694-A are $1.6\times
10^{-12}$ and $2.9\times10^{-13}\,{\rm erg \,cm}^{-2}{\rm s}^{-1}$, respectively.
The fluxes and line ratios are not corrected for extinction. No
aperture corrections have been applied.}
\end{table*}

\begin{table*}
\caption{Observed Line Ratios from the SH+LH spectral mapping data extracted
  with a $13.4\arcsec
  \times 13.4\arcsec$ aperture. }

\footnotesize

\begin{tabular}{lcccccccc}
\hline
Source & $f({\rm [NeII]}12.8\mu{\rm m})$ &
$\frac{ {\rm [NeIII]}15.6\mu{\rm m}}{{\rm [NeII]}12.8\mu{\rm m}}$ 
&  $\frac {{\rm [SIV]}10.5\mu{\rm m}} {{\rm [SIII]}18.7\mu{\rm m}}$  
&  $\frac {{\rm [SIII]}18.7\mu{\rm m}} {{\rm [NeII]}12.8\mu{\rm m}}$ 
& $\frac {{\rm H_2 S(2)}} {{\rm H_2 S(1)}}$ 
& $\frac {{\rm H_2 S(1)}}{{\rm [NeII]}12.8\mu{\rm m}}$ 
& $\frac {{\rm [SIII]}33.5\mu{\rm m}} {{\rm [SIII]}18.7\mu{\rm m}}$  
& $\frac {{\rm [SiII]}34.8\mu{\rm m}} {{\rm [SIII]}33.5\mu{\rm m}}$  \\
\\
\hline
A          & $30.5\pm 6.1$ &0.27 & 0.18 & 0.28 & 0.18 & 0.10 & 0.8 & 2.3\\
B1+B2      & $12.0\pm 2.4$ &0.52 & 0.22 & 0.40 & --   & -- & 1.2 & 1.4\\
C+C$^\prime$ & $18.8\pm 1.9$ &0.58 & 0.18 & 0.51 & -- & -- &1.0 & 1.2 \\
\hline
\end{tabular}

\tablecomments{
The [Ne\,{\sc ii}] fluxes are in units of 
$10^{-13}\,{\rm erg \,cm}^{-2}{\rm s}^{-1}$.
The fluxes and line ratios are not corrected for extinction. No
aperture corrections have been applied.}

\end{table*}

\section{The nuclear region of NGC~3690: the obscured AGN (B1) and surrounding
HII regions}

We start out the detailed analysis of Arp~299 with the nuclear region of
 NGC~3690. This area contains two bright $1.6\,\mu$m 
continuum sources, namely B1 and B2 (Fig.~5, lower panel). 
B2 is the optical nucleus, while B1 becomes the
 dominant  IR source at $\lambda > 2\,\mu$m in NGC~3690. B2 does not appear to
 be a bright source of mid-IR emission (Soifer et al. 2001).  
Direct evidence for an AGN in B1 
comes from hard X-ray observations  (Della Ceca et al. 2002; 
Zezas, Ward, \& Murray 2003; Ballo et al. 2004) and optical spectroscopy
(Garc\'{\i}a-Mar\'{\i}n et al. 2006). Given the large X-ray column
density ($N_{\rm H} = 2.5 \times 10^{24}\,{\rm cm}^{-2}$, Della Ceca et
al. 2002; Zezas et al. 2003)  toward this source the estimates of the X-ray
luminosity range from 
$L_{\rm X} = 1.1 \times 10^{41}\,{\rm erg \, s}^{-1}$ in the
$2-10\,$keV band (Ballo et al. 2004 from {\it XMM} observations), 
to $L_{\rm X} = 3-7 \times 10^{40}\,{\rm erg \, s}^{-1}$ in the 
$0.1-10\,$keV band (Zezas et
al. 2003 from {\it Chandra} data).
 Regardless of the uncertain absorption corrections for this
source, both estimates put it in the
 low-luminosity AGN category. Apart from the bright near-IR B1 and B2
 sources, there is star-formation around B1 
 detected both in  Pa$\alpha$   (see Fig.~5) and in the UV (Gallais
 et al. 2004 and Meurer et al. 1995),  indicating that these regions of star
 formation  must be
 relatively unobscured.

\subsection{AGN high excitation mid-IR emission lines}
The most direct way to identify an AGN in the mid-IR is through  high
excitation emission lines (e.g., Genzel et al. 1998;
Sturm et al. 2002; Mel\'endez et al. 2008). These high excitation lines 
are unlikely to be produced by star-formation, but they are not always
detected in relatively bright AGN (e.g., Weedman et al. 2005). 
Among the high excitation lines, the [Ne\,{\sc v}]$14.32\,\mu$m and [O\,{\sc 
    iv}]$25.89\,\mu$m lines are found to be the brightest in nearby AGN.
We note however, that extended [O\,{\sc 
    iv}]$25.89\,\mu$m line emission has been detected in some starburst
galaxies, indicating that it can also be associated with star formation (Lutz
et al. 1998). The [Ne\,{\sc v}]$14.32\,\mu$m
is not detected in the SH spectra of B1 and surrounding regions (see Fig.~6,
top panel).
From the SH spectrum  we placed a limit to the [Ne\,{\sc v}]$14.32\,\mu$m 
line of
$< 3 \times 10^{-14}\,{\rm erg \, cm}^{-2}\,{\rm s}^{-1}$, consistent
with the upper limit derived by Verma et al. (2003). 
The [O\,{\sc
    iv}]$25.89\,\mu$m line on the other hand is detected 
in the staring mode LH spectrum (bottom panel of Fig.~6), 
and we measured a
flux of   $ 2 \times 10^{-13}\,{\rm erg \,
    cm}^{-2}\,{\rm s}^{-1}$ (using an aperture correction for point source). 

It is illustrative to compare B1 in NGC~3690 with the ULIRG NGC~6240,
for which there is evidence for the presence of a double obscured AGN 
from X-ray observations (Komossa et al. 2003), radio observations (Gallimore
\& Beswick 2004),  and 
the detection of both [Ne\,{\sc v}]$14.32\,\mu$m and  [O\,{\sc
    iv}]$25.89\,\mu$m (Armus et al. 2006).
The ratios between the $0.2-10\,$keV (corrected for absorption) flux, and the 
[Ne\,{\sc v}]$14.32\,\mu$m and [O\,{\sc
    iv}]$25.89\,\mu$m line fluxes for NGC~6240 are approximately 20 and 4,
  respectively. 
Using the $0.1-10\,$keV flux from the {\it Chandra} high spatial resolution
data  for the B1 source of Zezas et al. (2003) and the above flux ratios, we
 predict line fluxes of 
$f([{\rm Ne\,V}])= 1-2 \times 
10^{-14}\,{\rm erg \, cm}^{-2}\,{\rm s}^{-1}$ 
and $f([{\rm O\,IV}])= 6-9 \times 
10^{-14}\,{\rm erg \, cm}^{-2}\,{\rm s}^{-1}$. The predicted 
[Ne\,{\sc v}] flux is in agreement with the
upper limit estimated above, whereas the predicted [O\,{\sc iv}] flux is below
the observed values.  
Similarly, using the correlation between the 
luminosities of the $2-10\,$keV emission and the 
[O\,{\sc    iv}]$25.89\,\mu$m line found by 
Mel\'endez et al. (2008) for a sample of Seyfert 1 galaxies, the estimated 
[O\,{\sc
    iv}]$25.89\,\mu$m
luminosity in B1 is slightly below our own measurement,
but within the scatter of the empirical correlation of Mel\'endez et
al. (2008). 

\begin{figure}

\includegraphics[width=6.cm,angle=-90]{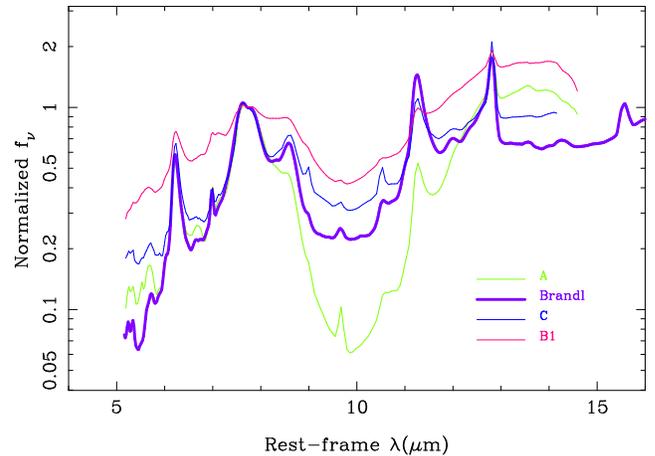}

\caption{SL spectra for a $3.7\arcsec \times 3.7\arcsec$ aperture 
of the three main sources of Arp~299 
compared to the average starburst spectrum of Brandl et al. (2006). All the
spectra are normalized at rest-frame  $7.7\,\mu$m.  }

\end{figure} 

\subsection{Hot dust emission in B1 and the composite nature of the mid-IR
  spectrum}

B1 appears as the brightest mid-IR continuum source of the
entire system out to $\lambda \sim 20\,\mu$m (Figs.~6 and 7, and also 
Soifer et al. 2001; Charmandaris,  Stacey, \& Gull 2002). 
Gallais et al. (2004) remarked
that this component shows a hot dust continuum at $5\,\mu$m, 
and they interpreted it as evidence for the
presence of an AGN. This hot dust component is 
even detected down to $2\,\mu$m (Shier et al. 1996; AAH00). 

The hot dust continuum  can be  
seen in the SL spectra of B1 in Fig.~8, where the 
spectra of the three brightest sources are normalized at the $7.7\,\mu$m 
PAH feature. To quantify the hot dust emission we used 
a diagnostic method similar to that of Nardini et al. (2008) 
to separate the AGN and starburst components using mid-IR spectra.
This method is based on the close similarity of the
$5 - 8\,\mu$m spectra of high metallicity starbursts (see Brandl et
al. 2006). Thus, any excess in the
$5-8\,\mu$m spectral region can be attributed to continuum emission from hot
dust. This hot dust component can be taken as an indicator of an 
AGN (see Nardini et al. 2008 for further details).  For our 
estimate we used two different templates, the  starburst template 
of Brandl et
al. (2006) and the star-forming ULIRG template of Nardini et
al. (2008). By using two different templates, one can 
estimate the uncertainties associated with this method.
We found that the  AGN contribution at $6\,\mu$m is 
$\sim 80-90\%$ (see Fig.~9)
for the $3.7\arcsec \times 3.7\arcsec$ spectrum. This example
illustrates the detection of a low-luminosity and/or heavily
obscured AGN  via the presence of a strong hot
dust continuum at relatively short mid-IR wavelengths ($3-6\,\mu$m,
see also Risaliti et al. 2006).

\begin{figure}

\includegraphics[width=9cm,angle=-90]{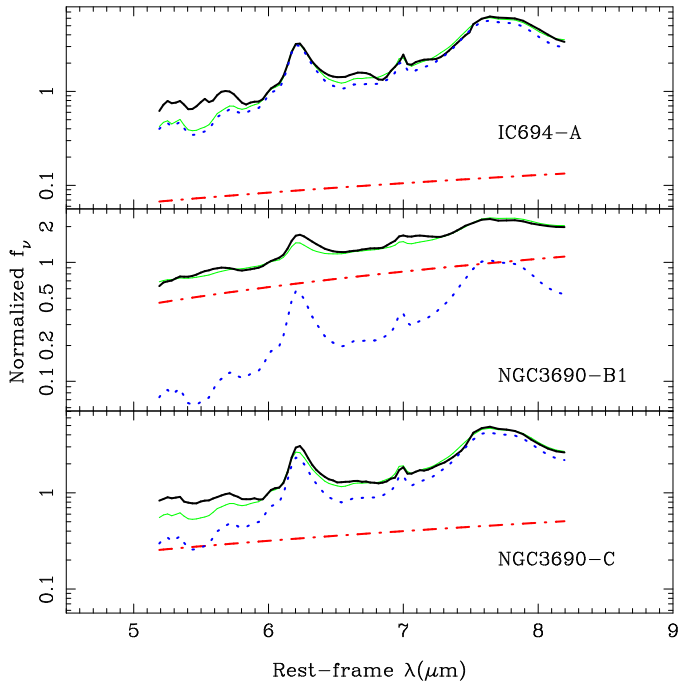}

\caption{Rest-frame $5-8\,\mu$m spectra (thick black lines)  
of the three main sources of Arp~299 observed with the SL1 module and 
extracted with $3.7\arcsec \times 3.7\arcsec$ apertures.
The observed spectra are normalized at $6\,\mu$m.
For each component we show as dot-dash red lines the fitted AGN continuum, as
dotted blue lines the fitted starburst component using the Brandl et al. (2006)
template, and as thin solid green lines the sum of the fitted AGN +
starburst components. We used a method similar to 
that of Nardini et al. (2008). 
}

\end{figure}

Given that the IRS apertures include the contribution from B1 (the AGN) and
the surrounding H\,{\sc ii} regions (see Fig.~5), it is reasonable to assume
that this region be identified as a composite source (AGN + SB) based
on several mid-IR diagnostics.  To measure the apparent 
strength of the $9.7\,\mu$m silicate feature, defined 
as $S_{\rm Si} = \ln f_{\rm obs} (9.7\mu{\rm m})/
f_{\rm cont}(9.7\mu{\rm m})$, 
we used a method similar 
to that outlined by Spoon et al. (2007)  for
sources dominated by PAH emission, fitting the continuum (see Fig.~8) 
as a power law with pivot wavelengths at 5.5 and $14\,\mu$m.  
Indeed, the measured EW of the $6.2\,\mu$m PAH feature ($\sim 0.2-0.3\,\mu$m) 
and the strength of the silicate
feature ($S_{\rm Si}=-0.80 \pm 0.1$) 
place source B1 in the Spoon et al. composite region, their ¨1B¨
region. Similarly, using the 
[Si\,{\sc ii}]$34.8\,\mu$m/[S\,{\sc iii}]$33.5\,\mu$m vs. 
EW($6.2\,\mu$m PAH) diagram from  Dale et al. (2006), or 
the [Ne\,{\sc v}]/[Ne\,{\sc ii}] vs. EW($6.2\,\mu$m PAH) diagram from 
Farrah et al. (2007), B1 would be classified as a composite system.

\section{The star-forming C+C$^\prime$ complex in the overlap region}

Gehrz et al. (1983) identified the C+C$^\prime$ 
complex in the overlap region of 
the two galaxies, 
as the most luminous starburst in Arp~299,
based on the observed H$\alpha$ luminosity. This region has been found to
be not only very luminous, but also one of the youngest star-forming 
complexes in the
system (e.g., Satyapal et al. 1999; Sugai et al. 1999; AAH00; 
Gallais et al. 2004; Garc\'{\i}a-Mar\'{\i}n et al. 2006), probably
as the result of  
the interaction process. High angular resolution imaging at different
wavelengths resolved the C+C$^\prime$ 
complex into a number of star clusters and  H\,{\sc ii} regions 
(Meurer et al. 1995; Aalto et al. 1997; Casoli et al. 1999; AAH00; 
Soifer et al. 2001).  C1, the brightest near-IR cluster (see AAH00 and
Fig.~10) in the C region, and C$^\prime$ are the 
most intense Pa$\alpha$ sources in this region (see Fig.~10).
Both appear to 
coincide with the bright radio and mid-IR sources in the overlap region.

The detection of the He\,{\sc i} $1.70\,\mu$m 
line in C (Vanzi, Alonso-Herrero, Rieke 1998; Sugai et al. 1999) 
indicates that the temperature of the ionizing stars is high ($T>40,000\,$K,
see Vanzi et al. 1996; F\"orster-Schreiber et al. 2001). The observed 
He\,{\sc i}$1.70\,\mu$m/Br10 line ratio (Sugai et al. 1999)
would imply an age younger than approximately 5\,Myr  for a solar metallicity
instantaneous burst (see models by Rigby \& Rieke 2004 and Snijders et. al
2007) or 7\,Myr for super-solar models. The weak near-IR CO
features (Ridgway, Wynn-Williams, \& Becklin 1994; Sugai et al. 1999; 
AAH00; Imanishi \& Nakanishi 2006) may indicate a lack of red supergiants
confirming that this region is relatively
young, but it is possible that the CO features are diluted by dust
emission (see \S6.3).

\subsection{Extinction and Physical Conditions}

The C+C$^\prime$  complex appears to undergo
varying degrees of obscuration based on the 
dust geometry as seen from the optical/near-IR
color map of AAH00 (their figure~9). The dust distribution appears to 
be  rather patchy and clumpy, especially near C1. This complicated
dust configuration probably contributes to 
the range of extinctions (from a few magnitudes up to $A_V=37\,$mag)
derived for this source using different optical and near-IR 
hydrogen recombination lines as well as different sized apertures (Satyapal et
al. 1999; Sugai et al. 1999; AAH00; 
Garc\'{\i}a-Mar\'{\i}n et al. 2006). 

\begin{figure}
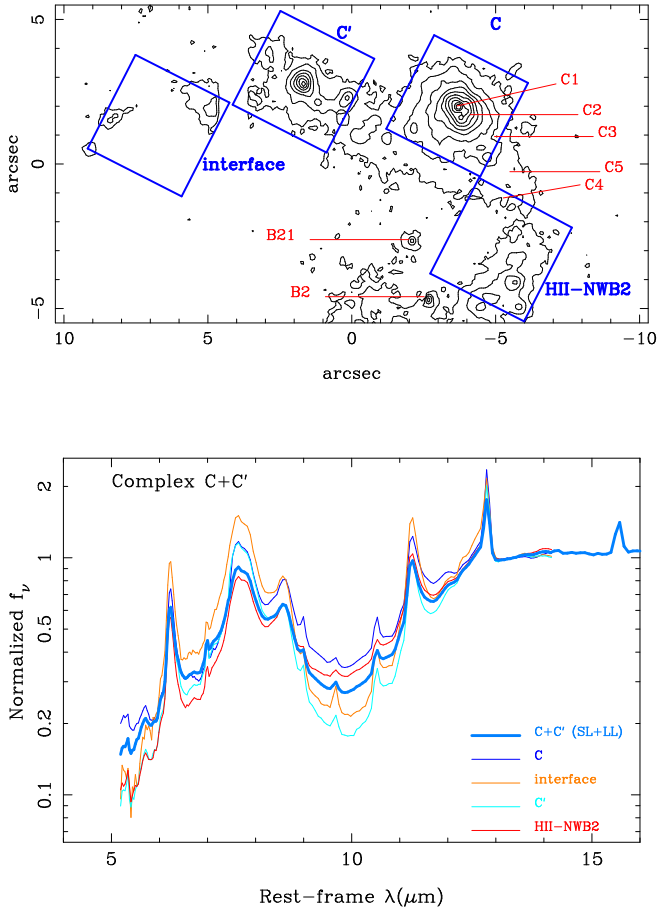


\includegraphics[width=5.cm,angle=-90]{f10a.ps}

\vspace{1cm}

\includegraphics[width=6.cm,angle=-90]{f10b.ps}

\caption{{\it Top:} Contours (linear scale) of the {\it HST}/NICMOS
  continuum-subtracted Pa$\alpha$ image 
of the C+C$^\prime$ complex region. The squares show the
  approximate locations of the SL1+SL2
  $3.8\arcsec \times 3.8\arcsec$ extraction apertures for four different
  regions. We also mark the positions of bright near-IR continuum 
star clusters in this
  region as well as the optical nucleus B2 
(see AAH00 for details). 
{\it Bottom:} SL1+SL2
  spectra of the regions shown in the top panel. We also 
show the $10.2\arcsec \times 10.2\arcsec$ SL+LL spectra 
that was fitted with {\sc pahfit}. All the spectra are normalized at rest-frame 
$14\,\mu$m, one of the pivot wavelengths used to measure the apparent depth of
  the $9.7\,\mu$m feature (see \S6.1).}

\end{figure}

The SL+LL spectrum of C+C$^\prime$ fitted 
with {\sc pahfit} with a dust screen model 
provides an optical depth of $\tau_{\rm si} = 0.8$ ($A_V \sim 13\,$mag) 
over a $10.2\arcsec \times 10.2\arcsec$ 
region. Since the  dust distribution of this region appears to be patchy, 
we used the
smallest possible SL1+SL2 apertures ($3.7\arcsec \times 3.7\arcsec
 \simeq 750\,{\rm pc} \times 750\,{\rm pc}$)  to 
extract spectra of four different regions in and around 
the C+C$^\prime$ complex, to better
isolate the different emitting sources. 
 Apart from sources C  and C$^\prime$, the other two  are a region  
at the interface of the galaxies southeast of C$^\prime$,
characterized by faint Pa$\alpha$ emission, and an H\,{\sc ii} region complex 
located approximately 3\arcsec
\,  northwest of B2. This H\,{\sc ii} region complex 
appears to be slightly older and less obscured than C and C$^\prime$, based on
optical data (see AAH00 and Garc\'{\i}a-Mar\'{\i}n et
 al. 2006). The approximate locations of the four regions are shown in Fig.~10.

We  measured the apparent strength of the $9.7\,\mu$m 
silicate feature from the four SL1+SL2 spectra as described in 
\S5.2.  We found $S_{\rm Si}=-0.44 \pm 0.05$, 
$-0.72 \pm 0.10$, $-0.54 \pm 0.05$, and $-0.25 \pm 0.05$, for C, 
C$^\prime$, the interface
region, and the H\,{\sc ii} region complex NW of B2, respectively. 
These translate into values of visual extinction of 
$A_V=7\pm 1\,$mag and $A_V=12\pm2\,$mag, to  C and C$^\prime$, 
respectively. These values might only be lower limits as
the broad PAH features at 7.7, 8.6 and $11.3\,\mu$m fill the
absorption feature resulting in an underestimate of the true optical
depth. For instance, using the SL+LL $10.2\arcsec \times 10.2\arcsec$ 
spectrum of the C+C$^\prime$ complex 
we obtained $S_{\rm Si}=-0.60 \pm 0.1$, whereas 
the optical depth fitted with {\sc
  pahfit} for the same region is $\tau_{\rm Si}=0.8$ (Table~2). In the case of
C, Sugai et al. (1999) using near-IR lines in the Brackett series derived 
$A_V\sim 8$mag  for a $4.6\arcsec$-diameter region, consistent with our
estimate.

The [S\,{\sc iii}]$33.48\,\mu$m/[S\,{\sc iii}]$18.71\,\mu$m line ratio can be 
used as an electron density estimator, as it has a small dependence on other
properties such as the
metallicity, ionization parameter, temperature of the stars, 
and the age of the ionizing stellar population
(see  e.g., Alexander et al. 1999; Snijders, Kewley, \& van
der Werf 2007). This line ratio is also most sensitive to densities 
in the range of $\sim 10^2- 10^4\,{\rm cm}^{-3}$. 
Tables~2 and 4 give this ratio as obtained from the low
(corrected for extinction) and
high (not corrected for extinction) 
spectral resolution data, respectively. 
The physical sizes covered by the
extraction apertures  are of the order of $2-2.7\,$kpc.

The [S\,{\sc iii}]$33.5\,\mu$m/[S\,{\sc iii}]$18.7\,\mu$m line 
ratios (Tables~2 and 4) of the C+C$^\prime$ complex are  
similar to those measured in   
other star-forming
galaxies on kiloparsec scales \\
(e.g., [S\,{\sc iii}]$33.48\,\mu$m/[S\,{\sc iii}]$18.71\,\mu{\rm m}
=1.2\pm0.3$ for the SINGS galaxies,  Dale et al. 2007) and
consistent with densities of 
the order of $n_{\rm e} \sim 300-500\,{\rm cm}^{-3}$ 
(see also Verma et al. 2003).

\subsection{Mid-IR Emission Lines and the Star-Formation Properties}

The number of ionizing photons can be estimated 
from the [Ne\,{\sc ii}] and [Ne\,{\sc iii}] line luminosities using 
expressions such as those given by Roussel et al. (2006) and 
Ho \& Keto (2007).  These relations depend 
mainly on the neon abundance, and the radiation temperature. The dependence
for the latter is through the fractional abundance of 
singly and doubly ionized neon. We took $T_{\rm rad}=4 \times 10^4\,$K. 
We also assumed that most of the neon in
Arp~299 is only singly and doubly ionized, consistent with young stars
being the dominant source of ionizing radiation. In contrast to Ho
\& Keto (2007), who used a fraction of 0.6 for the proportion of 
ionizing photons that are
absorbed by the gas as opposed to absorbed by dust or escaping,
we take this fraction to be equal to one,  
that is, $f_{\rm ion}=1$ in their equation~12. 
The last piece of  information needed  
to estimate the ionizing photons is the neon
abundance. Using {\it ISO} mid-IR spectroscopy 
Verma et al. (2003) measured super-solar neon abundances in the two
components of Arp~299.
Further evidence for super-solar abundances in this system comes from 
the oxygen abundance of $\log ({\rm O/H}) + 12 =8.92$ derived from  the
optical spectroscopy of
Moustakas \& Kennicutt (2006)  and
 using the Kobulnicky \& Kewley (2004) calibration
(John Moustakas, private communication). This oxygen abundance is
consistent with the range derived by Rupke, Veilleux, \& Baker (2008)
for a large sample of LIRGs in the local universe. 
We computed the number of ionizing photons using the expression given by
Ho \& Keto (2007) but using the neon abundances derived by Verma et
al. (2003) for NGC~3690, that is, [Ne/H]$=1.8\times 10^{-4}$ instead of the
solar value of   [Ne/H]$=9.9\times 10^{-5}$.

Under the assumptions stated above we  
find $N_{\rm Lyc} \sim 3\times 10^{53}\,{\rm s}^{-1}$ and  
$N_{\rm Lyc} \sim 2\times 10^{53}\,{\rm s}^{-1}$ for C and C$^\prime$,
respectively. The value for C is a factor of 10 lower than the
estimate from Pa$\alpha$ using $A_V=15$mag derived by AAH00. 
Using the values of $A_V=7$mag and $A_V=12$mag derived for C and 
C$^\prime$, and the
Br$\gamma$ fluxes of Sugai
et al. (1999), which were measured through $4.6\arcsec$-diameter apertures, 
we get 
$N_{\rm Lyc} \sim 2\times 10^{54}\,{\rm s}^{-1}$ and  
$N_{\rm Lyc} \sim 1\times 10^{54}\,{\rm s}^{-1}$,
respectively. The discrepancy between $N_{\rm Lyc}$
estimated from extinction-corrected Br$\gamma$ and from the neon fine
structure lines could be explained if some of the ionized gas is at densities
exceeding $n_{\rm crit} \sim 4 \times 10^5 {\rm cm}^{-3}$ 
for the neon lines. This
situation would arise if ultra-compact H\,{\sc ii} 
 regions play an important
role in the star formation in regions C and C', as suggested by Rigby \& Rieke
(2004) for luminous young starbursts in general. We return to this possibility
at the end of \S6.3.

The ionization parameter (defined as 
$q =N_{\rm Lyc}/(4\pi R^2n_{\rm
  ion})$) can then be estimated measuring approximate sizes of the nebulae 
from the {\it HST}/NICMOS Pa$\alpha$ images: 
$R \sim 300\,$pc and  $R\sim 250\,$pc, for C and C$^\prime$,
respectively, and taking $n_{\rm
  H}=300\,{\rm cm}^{-3}$. We find $q\sim 0.8-4 \times 10^8\,
{\rm cm\, s}^{-1}$  
for both C and C$^\prime$, using the neon line and hydrogen 
recombination line estimates of $N_{\rm Lyc}$.

We now make the simplifying
assumption that the ionized gas can be divided entirely between high and low
density regions. We use the mid-infrared fine structure lines to diagnose
conditions in the low-density region. 
Diagrams involving mid-IR line ratios of the same element can be
used to estimate the age of stellar population, although there is some
degeneracy between the age and the ionization parameter. 
We use the Snijders et al. (2007)  predictions 
of the age evolution of mid-IR fine-structure emission lines. 
These are calculated by combining the outputs of 
Starburst99    (Leitherer et al. 1999) and 
    the photoionization code Mappings (Dopita et al. 2000).
The stellar populations are assumed to
have been formed in an instantaneous burst of star formation with a
    Salpeter IMF (between 0.1 and $100\,{\rm M}_\odot$). 
Given the abundances measured in Arp~299 we used the $2Z_\odot$ models. 

Figure~11 is a  
[Ne\,{\sc iii}]$15.56\,\mu$m/[Ne\,{\sc ii}]$12.81\,\mu$m  
vs. [S\,{\sc iv}]$10.51\,\mu$m/[S\,{\sc iii}]$18.71\,\mu$m line ratio diagram
with the model predictions of Snijders et al. (2007). 
We show the models for two nebular densities, 
$n_{\rm H} = 10^2\,{\rm cm}^{-3}$, 
the low-density limit  typical of 
star-forming galaxies (see above), and 
$n_{\rm H} = 10^4\,{\rm cm}^{-3}$ to account for the high density nuclear
    region of IC~694 (see \S7.1). We also show the model predictions
    for three ionization parameters representative of those measured
    in different sources in Arp~299 (see also \S7.2 for A). 

The line ratios  (measured
    with the $4.5\arcsec \times 4.5\arcsec$ aperture) for C and C$^\prime$  
together with those of other sources\footnote{We also plotted in Fig.~11 the
  line ratios of B1. We caution the reader that with the spatial
  resolution of the IRS data  
we cannot estimate what fraction of the emission line fluxes 
originates from the AGN and what fraction 
from the surrounding H\,{\sc ii} regions. Thus we do not discuss the
star-formation properties of B1 based on this diagram} in the system are plotted in Fig.~11. 
The right panel gives the time evolution for the first few million
years. The line ratios drop very fast as the most
massive stars move off the main sequence with a minimum at about
3\,Myr.  After that the first Wolf-Rayet (W-R) stars
begin to appear the ratios increase again (left panel of Fig~11) for
the duration of the W-R phase (out to approximately 7\,Myr for the
super-solar models, Leitherer et al. 1999; Snijders et al. 2007).  
Given the measured ionization parameters for  C and C$^\prime$, 
we can rule out the youngest ages for these two sources. The mid-IR
line ratios in Fig.~11s could be reproduced with ages $4-7\,$Myr, using the 
$2Z_\odot$ models.  A summary of the mid-IR derived
 properties of the low density zones in 
C and C$^\prime$ sources is given in Table~5.
It is interesting to note however, that given the derived ranges of $q$ and
    ages, the [S\,{\sc iii}]$18.71\,\mu$m/[Ne\,{\sc ii}]$12.81\,\mu$m
line ratios can only be reproduced with relatively high nebular abundances 
($n_{\rm H}\ge 10^4\,{\rm cm}^{-3}$).

\begin{figure*}
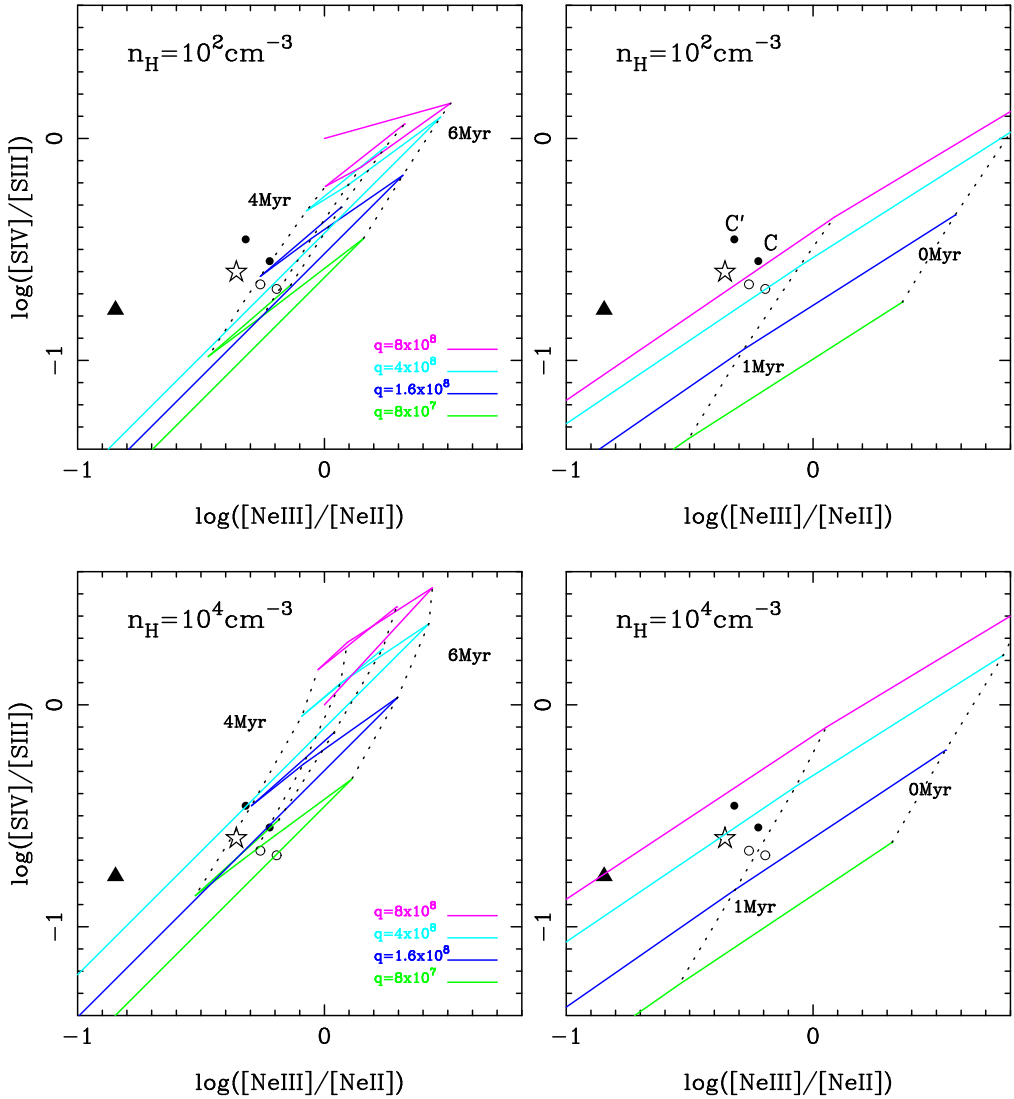


\hspace{2cm}
\includegraphics[width=7cm,angle=-90]{f11a.ps}

\vspace{0.5cm}

\hspace{2cm}
\includegraphics[width=7cm,angle=-90]{f11b.ps}

\caption{
Mid-IR line ratio diagrams. The symbols are as follows, 
the filled triangle is the extinction-corrected IC~694-A (using the 
scaling factors in table~2 of Farrah et al. 2007 for the Li \& Draine 2001
extinction law); the star symbol
is NGC~3690-B1; filled 
circles are the C and C$^\prime$ sources (marked for clarity in the upper right
panel); the open circles
are the F region in  IC~694, and the H\,{\sc ii} complex NW of B2 in NGC~3690. 
All the data are from the $4.5\arcsec \times
4.5\arcsec$ SH spectra. The solid lines are the Snijders et al. (2007)
models for $2Z_\odot$ and different ionization parameters. The top panels
are the $n_{\rm H} = 10^2\,{\rm cm}^{-3}$ models and lower panels 
are the $n_{\rm H} = 10^4\,{\rm cm}^{-3}$ models. The right panels show the
model evolution for 0, 1, and 2\,Myr, whereas the left panels are the model
evolution from 3.5, 4, 5, 6, and 7\,Myr. The dotted
lines connecting the different models represent the ``isochrones'' in these
  diagrams. }

\end{figure*}


\subsection{W-R features, hot dust and radio emission}

To further explore the possible presence of young stellar
populations in this region, we looked 
for the W-R broad He\,{\sc ii} feature at 4686\AA \ 
in the optical integral field
spectroscopy (IFS) of  Garc\'{\i}a-Mar\'{\i}n et al. (2006). 
The detection of this feature provides direct evidence
for the presence of a large population of hot, massive and young stars (Vacca
\& Conti 1992). The optical spectra of a few bright sources in Arp~299 are
shown in Fig.~12. The 
W-R broad He\,{\sc ii} feature at 4686\AA \  
is clearly detected in C and C$^\prime$, 
in the H\,{\sc ii} regions surrounding B1, and
tentatively detected in the H\,{\sc ii} region complex NW of B2. 
The high metallicity models predict more  W-R stars over an extended  
 lifetime compared with solar metallicity (Leitherer et
al. 1999). Specifically the $2Z_\odot$ models predict that  
stars enter the W-R phase at about  
2.5\,Myr and dominate the radiation
field out to 7\,Myr (see Snijders et al. 2007 for more details). 

C  and C$^\prime$ are also bright radio sources (Gehrz
et al. 1983; Neff et al. 2004) with observed radio spectral 
indexes\footnote{defined
as $S_\nu \propto \nu^\alpha$, where $S_\nu$ are the flux densities 
at 4.9 and 8.4\,GHz} of $\alpha=-0.65$ and $\alpha=-0.50$, respectively
(Neff et al. 2004). Models for the radio thermal emission of 
young stellar populations  ($<6\,$Myr) and solar metallicity predict 
$\alpha \ge -0.1$ (P\'erez-Olea \& Colina 1995), indicating that some 
fraction of the radio emission
must originate from supernovae. Using the $N_{\rm Lyc}$ derived in
\S6.2 from Br$\gamma$, the predicted thermal emission (see 
Colina \& P\' erez-Olea 1992) would account for approximately 
30\% of the
observed emission at 1.4\,GHz (Gehrz et al. 1983) from C through 
a $4\arcsec \times 4\arcsec $ aperture. We cannot make meaningful
comparisons at the other radio frequencies (4.9 and 8.4\,GHz), because
the measurements of Gehrz et al. (1983) were with beam sizes ($\sim
1\arcsec$) much smaller than our typical angular resolution (see \S2.1). 
Given this, it is possible that both C and C$^\prime$
are just now beginning a significant supernova activity, consistent
with the lack of young radio supernovae (Neff et
al. 2004). All these facts confirm
an age of C and C$^\prime$ of approximately $4-7\,$Myr  derived from
the mid-IR line ratios, 
rather than  extremely young bursts (see \S6.2).

Adding to this rather complicated situation,  is the presence of 
a hot dust component, which accounts for approximately $30-40$\% 
of the $6\,\mu$m emission in C 
within the $3.7\arcsec \times 3.7\arcsec$ SL aperture (see
Fig.~9).  A hot dust continuum has also been found based on the 
observed EW of
the 3.3$\mu$m PAH feature (Imanishi \& Nakanishi 2006). Hot dust is uncommon
in starbursts of the luminosity of C (Ridgway et al. 1994).
 However, it is unlikely that this dust is associated with a buried
AGN, as there is no other evidence for the presence of an
active nucleus.

The hot dust can be associated with ultra-compact 
regions, whose presence is also
indicated by the discrepancy in $N_{\rm Lyc}$ 
deduced from Pa$\alpha$ and the neon fine
structure lines. It is likely that what we are calling source C is in
reality a collection of different sources with different ages, 
as already hinted by the
detection of a number of near-IR continuum sources in the large 
Pa$\alpha$ nebula (see Figs.~5 and 10). Source C is reminiscent of the nuclear 
region of the W-R galaxy Henize 2-10, which hosts young clusters,
ultra-dense H\,{\sc ii} regions, and possibly supernova remnants
within the central few hundred parsecs (see Cabanac, Vanzi, \&
Sauvage 2005; Mart\'{\i}n-Hern\'andez et al. 
2006). 

The discovery that ultra-compact regions play
an important role in the young star-forming regions in C and C' supports the
hypothesis of Rigby \& Rieke (2004). They proposed that large numbers of such
regions would account for the lack of high-excitation emission lines in
starbursts without imposing restrictions on the initial mass functions in
them.

\begin{figure}
\includegraphics[width=9cm,angle=0]{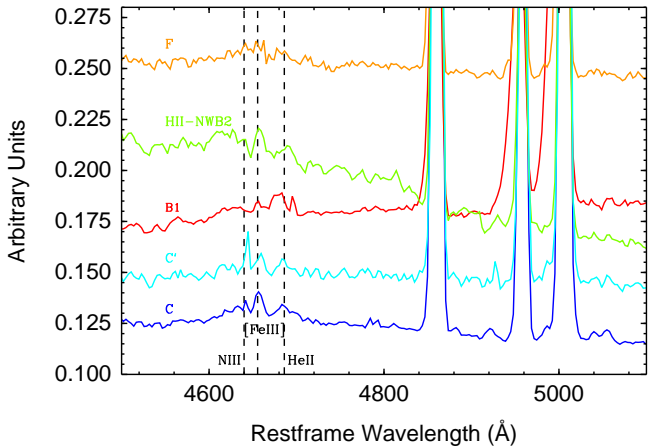}

\caption{Optical spectrum of a number of sources in Arp~299 
from the IFS of Garc\'{\i}a-Mar\'{\i}n
  et al. (2006) extracted with their large 
fiber of 3\arcsec \, ($\sim 610\,$pc) in 
diameter to match as much as possible the small {\it Spitzer}
apertures. 
The W-R broad He\,{\sc ii} feature at 4686\AA \  
is clearly detected in C and C$^\prime$, 
and marginally detected in the H\,{\sc ii} region
complex NW of B2. }

\end{figure}

\section{The nuclear region of IC~694 (A)}

From the early studies of this interacting system, the star-formation
properties of the nuclear region of IC~694 were found difficult to
model, in part due to the strong dust extinction that obscures most of the
H$\alpha$ emission and in part due to  the presence of 
bright radio emission (Gehrz et al. 1983). It also became 
 apparent that this source alone 
might be responsible for about half of the total IR
luminosity of the system (Gehrz et al. 1983; Joy et al. 1989; 
Keto et al. 1997; Charmandaris et al. 2002). 
Near-IR imaging and spectroscopic 
observations were able to partially overcome the strong obscuration to the
nucleus of IC~694. Sugai et al. (1999) and  
AAH00 showed that this nuclear region is affected by strong obscuration, and
that the observed properties are the result of an extended period of star
formation (see also Shier et al. 1996). 
In fact, the star-formation activity of 
source A appears to be older than that of 
the other bright components of the system. This
conclusion was later reinforced by the detection of a number of 
radio sources in the central 0.6\arcsec \  of A, which could 
only be explained as supernova remnants (Neff et al. 
2004; Ulvestad, Johnson, \& Neff 2007) from  a relatively evolved 
star formation ($6-8\,$Myr after the peak of the activity). 
Other works suggested the presence of
an obscured AGN in the nucleus of IC~694. This possibility is discussed in
\S7.3. 

\begin{table*}
\begin{center}

\caption{Summary of Mid-IR derived properties of the main components of 
Arp~299}
\begin{tabular}{lcccccc}
\hline
Source & $A_V$ & $n_{\rm e}$ & Age & $N_{\rm Lyc}$ & $q$ & Hot dust\\
(1)    & (2)   & (3) & (4) & (5) & (6) & (7)\\
\hline
A        & 34 & $1000-5000$ & $>7$ & $8$ & $1-5\times10^{8}$& 10-20\%\\
B1       & 14 & $250-400$ & --- & --- & --- & 80-90\% \\
C        & 7  & $250-400$ & $4-7$ & $3$ & $0.8-4\times10^{8}$&  30-40\% \\
C$^\prime$ & 12 & $250-400$ & $4-7$ & $2$ & $0.8-4\times10^{8}$&  -- \\
\hline
\end{tabular}

Notes.---- Column~(1): Source. Column~(2): Optical extinction in magnitudes
as derived from the optical depth of the $9.7\,\mu$m absorption feature. 
For A the value is that obtained from {\sc pahfit}, whereas for the other
sources the apparent depth was measured as explained in \S5.2.
Column~(3): Electron density in cm$^{-3}$ from the 
[S\,{\sc iii}]$33.5\,\mu$m/[S\,{\sc iii}]$18.7\,\mu$m line ratio. 
Column~(4): Age estimate in Myr from mid-IR line ratios. 
Column~(5): Number of ionizing photons in units of 
$10^{53}\,{\rm s}^{-1}$ estimated from 
the [Ne\,{\sc ii}]$12.81\,\mu$m and [Ne\,{\sc iii}]$15.56\,\mu$m emission
lines. The two neon line fluxes were measured in $4.5\arcsec \times
4.5\arcsec$ apertures for C and C$^\prime$, 
and for  $6.8\arcsec \times
6.8\arcsec$ for source A. The neon abundances used 
are 2.2 and 1.7 solar for A, and C+C$^\prime$, respectively (Verma et
al. 2003). Column~(6): 
Range of possible ionization parameters in units of cm s$^{-1}$, using
the two estimates of the number of ionizing photons, from neon lines
and Br$\gamma$ measurements of Sugai et al. (1999). 
Column~(7): Hot dust contribution to
the $6\,\mu$m luminosity using a  method similar to that described by Nardini 
et al. (2008).  

\end{center}
\end{table*}

\subsection{Extinction and Physical Conditions}

The overall shape of the 
low-resolution spectra of the nuclear region of IC~694 appears to be
remarkably different from the other components of Arp~299 (see Fig.~7 and 8,
and Gehrz et al. (1983). 
The mid-IR spectrum of A shows a  much deeper 
absorption at $9.7\,\mu$m (see also Gallais et al. 2004), 
and $18\,\mu$m than the other components in the system 
and  than the average
spectrum of starburst galaxies (Brandl et al. 2006).  The depth of the
silicate feature is comparable to
the values  observed in local ULIRGs (see e.g., Spoon et al. 2007; 
Armus et al. 2007; Farrah et al. 2007). 
{\sc pahfit} fits  a $9.7\,\mu$m apparent 
optical depth of $\tau_{\rm si} = 2.04$  
for the dust screen model. Using the Rieke \& Lebofsky (1986)
extinction law this optical depth implies an extinction of $A_V 
\sim  34\,\,$mag (see also Gallais et al. 2004), 
 comparable to, although higher than, the near-IR 
estimates (e.g., Beck, Turner, \& Ho 1986; Sugai et al. 1999; AAH00).  
We note, however, that the observed apparent optical depth of A and
other ULIRGs cannot be reproduced with a foreground dust screen, and
they require the dust to be in a deeply embedded and optically thick
smooth distribution (Levenson et al. 2007).

The nuclear region of IC~694 over a 
2\,kpc scale shows a
much lower [S\,{\sc iii}]$33.48\,\mu$m/[S\,{\sc iii}]$18.71\,\mu$m line 
ratio (corrected for extinction) 
than the nuclear region of NGC~3690 
(B1) and the C+C$^\prime$ complex. 
Using the photoionization models of Snijders et al. (2007),
the observed sulfur ratios of the central $2-2.7\,$kpc  
of IC~694 would imply densities of the order of $n_{\rm e} \sim
1- 5\times 10^3\,{\rm cm}^{-3}$ (see also models of Alexander et
al. 1999). This indicates that the nuclear region of IC~694 has a density 
higher  than the typical values inferred for starburst galaxies
from mid-IR spectra ($n_{\rm e} \sim 300-400\,{\rm cm}^{-3}$, Verma et al.  
2003; Dale  et al. 2006).

The high density of the nuclear region of IC~694 derived from the 
[S\,{\sc iii}] lines is in agreement
with the modelling of the radio H92$\alpha$ transition of Zhao et al. (1997).
They found values of  
$n_{\rm e} \sim 5\times 10^2-5\times10^4\,{\rm cm}^{-3}$ at the very center
and $n_{\rm e} \sim 5\times 10^3-10^4\,{\rm cm}^{-3}$ in the nuclear extension
about $2\arcsec$ southeast of the center (also seen in the {\it HST}/NICMOS  
Pa$\alpha$ map, Fig.~5).  Aalto et al. (1997) also found that the  
high $^{12}$CO/$^{13}$CO molecular line ratio of the A nuclear region implied
a population of unusually dense and warm clouds, due primary to small or
moderate optical depth of the 1-0 $^{12}$CO transition.

\subsection{Mid-IR Emission Lines and the Star-Formation Properties}

The number of ionizing photons arising from the central region of IC~694 has
been estimated by a number of IR and radio 
methods, and is in the range $N_{\rm Lyc} \sim 1-4
\times 10^{54}\,{\rm s}^{-1}$ (Shier, Rieke, \& Rieke 1996; Zhao et al. 1997;
Sugai et al. 1999; AAH00). We can obtain an independent
estimate 
from the extinction-corrected\footnote{$A_V=34\,$mag and the Rieke \&
Lebofsky (1985) extinction law}
[Ne\,{\sc ii}]$12.81\,\mu$m and [Ne\,{\sc iii}]$15.56\,\mu$m  luminosities 
measured with the 
SH $6.8\arcsec \times 6.8\arcsec$ aperture, as explained
in \S6.2 and using  the neon abundance inferred by 
Verma et al. (2003).  We find $N_{\rm Lyc} \sim 
0.8 \times 10^{54}\,{\rm s}^{-1}$ from the neon lines. As found for C
and C$^\prime$ the number of ionizing photons from the neon lines is a
factor of a few lower than other estimates from hydrogen lines. 
The corresponding range of ionization parameters 
is $ q \sim 1-5\times 10^8\,{\rm cm \,s}^{-1}$
using a size of the nebula of $R\sim 200\,$pc measured 
from the {\it HST}/NICMOS Pa$\alpha$ image), 
$n_{\rm H} = 1000\,{\rm cm}^{-3}$, and the number of ionizing photons
from the neon lines and the Br$\gamma$ measurements of Sugai et al. (1999).

The observed mid-IR line ratios 
([Ne\,{\sc iii}]$15.56\,\mu$m/[Ne\,{\sc ii}]$12.81\,\mu$m  
and [S\,{\sc iv}]$10.51\,\mu$m/[S\,{\sc iii}]$18.71\,\mu$m)
of source A are shown in Fig.~11 with the filled triangle 
symbol.  Since the highest 
possible ionization parameter for A would be 
$q\sim 4 \times 10^8\,{\rm cm \,s}^{-1}$ (i.e., using the number of ionizing
photons from the Br$\gamma$ line of Sugai et al. 1999), 
we can rule out the youngest
age for A ($1-2\,$Myr, lower right panel of Fig.~11). 
The mid-IR line ratios from A would be better explained with an 
evolved starburst $>6-7\,$Myr (lower left panel of Fig.~11). 
This is consistent with the age derived from the radio and near-IR
properties. We also note that models with constant
or quasi-constant  (e.g., exponentially decaying with long 
timescales) star formation
can also be excluded for this source as such models 
would predict high mid-IR line ratios for all
ages (see Thornley et al. 2000; Rigby \& Rieke 2004).

\subsection{No mid-IR evidence for an obscured AGN}
Early suggestions for the presence of an obscured AGN in A were based on
the compact nature of the radio emission (e.g., Sargent \& Scoville 1991;
Lonsdale, Lonsdale, \& Smith 1992). Ballo et al. (2004)  measured an X-ray
luminosity for A similar to that of B1 (see \S5), and found 
it could not be accounted
for by star formation alone. In contrast,
Zezas et al. (2003) were able to explain this luminosity in terms of high mass
X-ray binaries as estimated from the number of ionizing photons. The 
water megamaser properties of this nucleus do not exclude the possibility of 
an AGN in A (Tarchi et al. 2006).

As for B1 we did not detect the high excitation [Ne\,{\sc v}]$14.32\,\mu$m 
emission line, typical
of AGN. The [O\,{\sc iv}]$25.89\,\mu$m is detected in the LH staring mode
spectrum (bottom panel of Fig.~6) with a measured flux of $1\times
10^{-14}\,{\rm erg\,cm}^{-2}\,{\rm s}^{-1}$. 
Unfortunately the limited spatial resolution of the LH data does not allow us
to determine whether the integrated [O\,{\sc iv}] emission overr the 
LH slit ($\sim
11\arcsec \times 22\arcsec$) is extended or not. 
Using the rest-frame $5-8\,\mu$m spectrum and the
 method described in \S5.2,
the contribution from hot dust at $6\,\mu$m is  estimated to be 
approximately 20\%. Given the uncertainties of the method, 
this contribution is formally compatible with pure star formation 
(see Fig.~9). 
Finally the location of A on the diagram of EW of the $6.2\,\mu$m PAH feature
vs the depth of the silicate feature (Spoon et al. 2007) is similar to that of
M82, and thus consistent with obscured star formation.

\begin{figure}
\includegraphics[width=6.3cm,angle=-90]{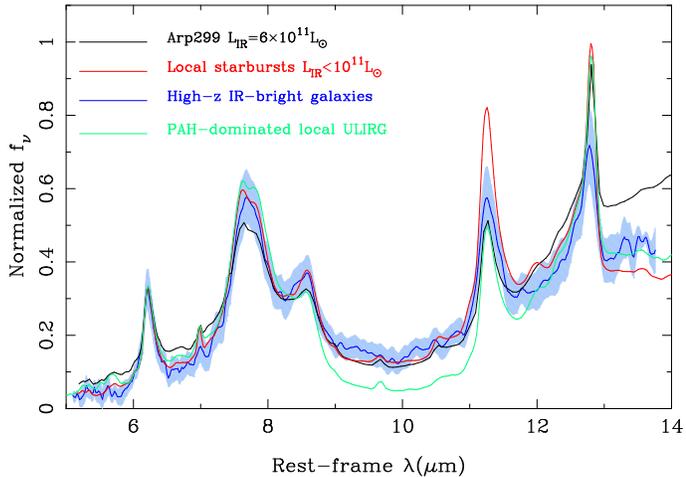}

\caption{Comparison of the low spectral-resolution spectra of the integrated
  emission of Arp~299 with the average starburst template of Brandl et
  al. (2006), the average spectrum of PAH-dominated local ULIRGs (the
  2C class of Spoon et al. 2007), and the average spectrum and corresponding 
$1\sigma$ dispersion of high-$z$ ULIRGs 
from the Farrah et al. (2008) sample. The IR ($1-1000\,\mu$m) 
luminosity range of the 
galaxies included in this high-$z$ template is $\sim 8 \times 10^{12}$ to 
$6 \times 10^{13}\,{\rm L}_\odot$. The
  spectra have been scaled to match the peak of the $6.2\,\mu$m PAH feature.}

\end{figure} 

\section{Arp~299 in the context of local and high-z 
IR-bright galaxies}

The unprecedented sensitivity of IRS on {\it Spitzer} is making it
possible to characterize the mid-IR spectra of IR bright-galaxies
at $z>1$. Perhaps one of the most surprising results is that
a large fraction of these high-$z$ IR-selected galaxies are classified
as ULIRGs in terms of their IR luminosities,
but their mid-IR spectra are more similar to those of local
star-forming galaxies (e.g., Yan et al. 2007; Farrah et
al. 2008; Rigby et al. 2008). That is, ULIRGs at high-$z$ show
strong PAH features and moderate depths of
the $9.7\,\mu$m silicate feature. This is in contrast with 
their local counterparts that  
tend to show very deep silicate features and moderate to 
low equivalent widths of
the PAHs (e.g., Spoon et al. 2006; Armus et al. 2007; 
Farrah et al. 2007).

Figure~13 shows a comparison of the integrated spectrum of Arp~299 with the
average template of starbursts from Brandl et al. (2006). It is
immediately clear 
from this figure that the two mid-IR $5-14\,\mu$m spectra are quite
similar, at least in terms of the PAH features, 
despite the different range of IR luminosities of the
galaxies included in the starburst template\footnote{This template is the
  average of 13 starburst galaxies,  of which
two galaxies are just  above the LIRG limit, whereas the rest of the galaxies 
have IR luminosities between $6\times 10^9\,{\rm L}_\odot$ and $6\times
10^{10}\,{\rm L}_\odot$.}.   
It is also clear that the integrated mid-IR 
spectrum of Arp299 is not dominated
by the strongly obscured source in the nuclear region of IC~694 (see
Fig.~7). The apparent strength of the $9.7\,\mu$m silicate feature of the
integrated spectrum is $S_{\rm Si}=-0.79 \pm 0.05$, 
similar to the typical values shown by starburst galaxies (Spoon et
al. 2007; Brandl et al. 2006), whereas for source A
we find $S_{\rm Si}=-2.03 \pm 0.10$. We also show in this comparison
the average template of local PAH-dominated ULIRGs (the 2C class of
Spoon et al. 2007), which is similar in terms of the PAH features but
it has a deeper silicate
feature. The other dominant class of local ULIRGs (not shown 
in Fig.~13), the 3C class in the
Spoon et al. (2007) classification, has a much deeper silicate feature
and  low EW of the PAH features, and it is
even more different from the integrated spectrum of Arp~299 .

We also show in Fig.~13 the average spectrum of the sample of high-$z$
ULIRGs (and even  higher IR luminosities) 
of Farrah et al. (2008). The galaxies in this  sample were selected such that
they showed a  strong
$1.6\,\mu$m feature and bright $24\,\mu$m fluxes, and are mostly in a narrow
redshift range around $z=1.71$.
The similarity between the integrated
spectrum of Arp~299 and these high-$z$ ULIRGs is quite remarkable. 
One of the favored explanations for the observed mid-IR spectra of high-$z$
ULIRGs is that star formation is diffuse and extended,  
or distributed in multiple dusty star-forming regions spread over
 several kiloparsecs. Arp~299 behaves in this manner and may be a 
local counterpart of the star formation processes taking place in 
high-$z$ IR-bright galaxies (see also
 Charmandaris,  Le Floc´h, \& Mirabel 2004), albeit with a lower IR
 luminosity and possibly higher metallicity. This is in contrast with  
local ULIRGs where most of 
the star-formation is occurring in the nuclei of the galaxies on 
sub-kiloparsec scales (Soifer et al. 2000).

\section{Conclusions}
We have presented {\it Spitzer}/IRS spectral 
mapping of the central $\sim 45\arcsec \sim 9\,{\rm kpc}$ 
of Arp~299 (IC~694 + NGC~3690). 
The IRS data combined with {\it HST}/NICMOS imaging 
and optical ground-based IFS 
allowed us to study in detail the
morphological distribution of mid-IR spectral features as well
as the mid-IR properties of the brightest sources in the system. 
We used a number of mid-IR indicators 
to derive the star formation properties
(ages, number of ionization  photons, and ionization parameters) 
and physical conditions 
of selected regions in the system. The main 
conclusions of this study are:

\begin{itemize}

\item
The spatial distribution of the bright mid-IR emission lines 
follows well the ionizing stellar populations as probed by the {\it
    HST}/NICMOS Pa$\alpha$ emission. High  [Ne\,{\sc
    iii}]$15.56\,\mu$m/[Ne\,{\sc ii}]$12.81\,\mu$m and 
[S\,{\sc iv}]$10.51\,\mu$m/[S\,{\sc
    iii}]$18.71\,\mu$m line ratios mark  the location of the youngest 
regions, that is, the C+C$^\prime$ complex and H\,{\sc ii} regions in
the disks of the galaxies. 
\item
Most regions in the system show 
[S\,{\sc iii}]$33.48\,\mu$m/[S\,{\sc iii}]$18.71\,\mu$m line
ratios similar to those typical of star-forming galaxies, 
implying densities of the order of
$n_{\rm e} \sim 250-500\,{\rm cm}^{-3}$. Only the ratios of the 
nuclear region of IC~694 indicate  higher densities, 
$n_{\rm e} \sim 1000-5000\,{\rm cm}^{-3}$, in agreement with previous 
results for this source 
from radio hydrogen emission and molecular CO line ratios.

\item
The extinction as estimated from the optical depth of the $9.7\,\mu$m silicate
feature is found to vary significantly from region to region in Arp~299. For 
the nuclear region of IC~694 we measured $\tau_{\rm Si}=2$ or $A_V=34\,$mag,
comparable to the optical depths measured in local ULIRGs.  The other bright
sources of the system (B1, C, and C$^\prime$) 
show more moderate values of the extinction ranging from
7 to 14\,mag.
\item
Both $6.2\,\mu$m and  $11.3\,\mu$m PAH emission is detected in the bright
sources of the system as well as at the interface region connecting  the two galaxies. There is a tendency for the observed $11.3$ to $6.2\,\mu$m PAH
feature ratio to be higher in regions of low ionization, although the observed 
$11.3\,\mu$m PAH feature emission can be significantly 
reduced in regions with elevated extinction. 

\item
The mid-IR line ratios suggest that youngest stellar 
populations ($\sim 4-7\,$Myr, using $2Z_\odot$ models) are mainly 
located in the
overlap region of the two galaxies (the C+C$^\prime$ complex), 
the southern spiral 
arm of IC~694, and the H\,{\sc ii} region complex to the northwest of the
nuclear region of NGC~3690. The detection of optical W-R features in C and
C$^\prime$ provides further evidence for the youth of these
regions. The extended lifetimes of super-solar metallicity W-R stars
means that they can
dominate the radiation field to older ages (out to $\sim 7\,$Myr). In
contrast, the mid-IR properties
of the nuclear region of IC~694, in particular the low [Ne\,{\sc
  iii}]/[Ne\,{\sc ii}] line ratios, 
confirm that the star formation activity there 
is slightly older and more extended
    in time than in the aforementioned regions.
\item
Evidence for hot dust emission in the  $5-8\,\mu$m 
spectral region for B1, the nuclear region of NGC~3690. 
B1 hosts an X-ray and optically identified
low-luminosity AGN. The AGN is likely to be 
responsible  for $\sim 80-90\%$ of the  $6\,\mu$m emission (within a
$3.7\arcsec  \times 3.7\arcsec$ aperture). 
\item
In C, the overlap region between the two galaxies and a site of very recent
star formation, we found both hot dust and also evidence that a 
portion of the ionized gas may be in very dense regions, above the critical
densities for [Ne\,{\sc ii}] and [Ne\,{\sc iii}]. The hot dust continuum
contributes $\sim 30-40\%$ at $6\,\mu$m. Given that there is no other
indication for the presence of an AGN in C,  we interpreted it as
produced by  dust heated by a young star cluster or clusters. 
Both of these results may be 
consistent with the presence of a substantial population of ultra-compact 
H\,{\sc ii} regions.  
\item
The integrated mid-IR spectrum of 
Arp~299 is  similar to that of local starbursts 
despite  its strongly interacting nature and high IR luminosity.
This is because the star formation in this system is spread 
across at least 6-8\,kpc, with a large fraction taking place 
in regions of moderate mid-IR optical
depths. It is only the nuclear region of IC~694 that shows the typical mid-IR
characteristics of ULIRGs, that is, very compact and embedded star
formation resulting in a deep silicate feature and
moderate EW of the PAHs. We also find that the integrated 
$5-14\,\mu$m spectrum of Arp~299 is similar to the average spectrum of 
the high-$z$ sample of Farrah et al. (2008). This suggests that the 
Arp~299 system 
may represent a local example, albeit with lower
IR luminosity and possibly higher metallicity, of the processes
occurring in high-$z$ IR-bright
galaxies. 

\end{itemize}

\acknowledgements

We would like to thank the  referee for useful comments that improved the
paper.  The authors would also like to thank Guido Risaliti, 
Henrik Spoon, John Moustakas, and Duncan
Farrah  for very their help and enlightening discussions, as well as 
to Tanio D\'{\i}az-Santos for constructing the AORs for the
spectral mapping and Miwa Block for producing the data cubes. 
 This work was supported by NASA through contract 1255094 issued by
 JPL/California Institute of Technology.  AA-H, LC and MP-S acknowledge 
support from the Spanish Plan Nacional del Espacio under grants ESP2005-01480
and ESP2007-65475-C02-01.
MG-M is supported by the German federal department for education and
research (BMBF) under the project numbers: 50OS0502 \& 50OS0801.
This research has made use of the NASA/IPAC Extragalactic Database (NED),
 which is operated by the Jet Propulsion Laboratory, California Institute of
 Technology, under contract with the National Aeronautics and Space
 Administration.

\end{document}